\documentclass[]{elsarticle}

\usepackage[margin=2.5cm]{geometry}

\usepackage{booktabs}
\usepackage{lineno}
\usepackage{amsmath}
\usepackage{amssymb}
\usepackage{xcolor}
\usepackage{caption}
\usepackage{subcaption}
\usepackage{rotating}
\usepackage{listings}
\usepackage{algorithm}
\usepackage[noend]{algpseudocode}

\usepackage{float}

\graphicspath{{pics/}}

\journal{-}

\begin{document}

\begin{frontmatter}

\title{HRRRCast: a data-driven emulator for regional weather forecasting at convection allowing scales}

\author[cires,gsl]{Daniel Abdi\corref{equal}}
\author[gsl]{Isidora Jankov\corref{equal}}
\author[cires,gsl]{Paul Madden}
\author[cira,gsl]{Vanderlei Vargas}
\author[psl]{Timothy A. Smith}
\author[psl]{Sergey Frolov}
\author[okl,nssl]{Montgomery Flora}
\author[nssl]{Corey Potvin}

\address[cires]{Cooperative Institute for Research in Environmental Sciences (CIRES), CU, Boulder, CO}
\address[cira]{Cooperative Institute for Research in the Atmosphere (CIRA), CSU, Fort Collins, CO}
\address[okl]{Cooperative Institute for Severe and High-Impact Weather Research and Operations (CIWRO), OU, Norman, OK}
\address[gsl]{NOAA Global Systems Laboratory, Boulder, CO}
\address[psl]{NOAA Physical Sciences Laboratory, Boulder, CO}
\address[nssl]{NOAA National Severe Storms Laboratory, Norman, OK}

\cortext[equal]{These authors have made equal contributions.}

\begin{abstract}
The High-Resolution Rapid Refresh (HRRR) model is a convection-allowing model used in operational weather forecasting across the contiguous United States (CONUS). To provide a computationally efficient alternative, we introduce HRRRCast, a data-driven emulator built with advanced machine learning techniques. HRRRCast includes two architectures: a ResNet-based model (ResHRRR) and a Graph Neural Network-based model (GraphHRRR). ResHRRR utilizes convolutional neural networks enhanced with squeeze-and-excitation blocks and Feature-wise Linear Modulation, and supports probabilistic forecasting via the Denoising Diffusion Implicit Model (DDIM). To better handle longer lead times, we train a single model to predict multiple lead times (1h, 3h, and 6h), and then use a greedy rollout strategy during inference. When evaluated on composite reflectivity over the full CONUS domain using ensembles of 3 to 10 members, ResHRRR outperforms HRRR forecast at light rainfall threshold (20 dBZ) and achieves competitive performance at moderate thresholds (30 dBZ). Our work advances the pioneering StormCast model described in \citet{pathak2024} by:
a) training on the full CONUS domain,
b) training on multiple lead times to improve long-range performance,
c) using analysis data for training instead of the +1h post-analysis data inadvertently used in StormCast, and
d) incorporating future Global Forecast System (GFS) weather states as inputs, adding a downscaling component that significantly improves long-lead forecast accuracy.
Grid-based, neighborhood-based, and object-based verification metrics confirm improved storm placement, lower frequency bias, and enhanced success ratios compared to HRRR. Additionally, HRRRCast's ensemble forecasts maintain sharper spatial detail and reduced blurriness than deterministic baselines, with power spectra more closely matching HRRR analysis. While GraphHRRR underperforms in its current form, it lays the groundwork for future probabilistic graph-based forecasting. Overall, HRRRCast represents a step toward efficient, data-driven regional weather prediction with competitive accuracy and ensemble capability.
\end{abstract}


\end{frontmatter}



\section{Introduction}
Recent advances in machine learning weather prediction (MLWP) have shown great promise in complementing or even replacing traditional numerical weather prediction (NWP) systems, particularly at global scales. Several studies have demonstrated that data-driven models can rival the skill of physics-based models at a fraction of the computational cost, enabling applications such as ensemble forecasting and climate downscaling with greater efficiency \citep{bi2023,keisler2022,lam2023,price2024,nguyen2023,nguyen2023b}.

However, while progress in global MLWP is substantial, the transition to high-resolution regional forecasting--especially at convection-allowing scales (km-scale) -- remains an active area of research. Regional forecasting presents unique challenges: the atmospheric dynamics at these scales are strongly non-hydrostatic and involve rapid, stochastic convective processes that are difficult to emulate with deterministic models \citep{pathak2024,larsson2025}. Furthermore, regional models must resolve localized phenomena such as thunderstorms, mesoscale convective systems, and orographic precipitation, which are often poorly captured by coarser global models.

Among several recent efforts addressing these challenges\citep{Flora2025,oskarsson2023,pathak2024,adamov2025,larsson2025}, the StormCast model \citep{pathak2024} is notable for introducing a generative diffusion-based approach to emulate the High-Resolution Rapid Refresh (HRRR) \citep{dowell2022} model at 3 km resolution. By autoregressively predicting full 3D atmospheric states conditioned on synoptic-scale global forecasts, StormCast achieved skill in predicting convective structures such as composite reflectivity out to 6h lead time. 

In this work, we present HRRRCast, a  data-driven emulator for regional weather forecasting that builds upon and extends the capabilities of StormCast and related efforts \citep{pathak2024,oskarsson2023,adamov2025}. Our model aims to approximately emulate the HRRR across the CONUS, albeit at a coarser 6 km grid spacing, and addresses several limitations:

\begin{itemize}
\item \textit{Full CONUS domain:} Unlike StormCast, which focused on a regional subset at 3km resolution, our model is trained on the entire CONUS domain at 6km resolution. This enables full forecast coverage, and enhances the model's generalizability by exposing it to a wider range of convective weather patterns and meteorological regimes across the CONUS.
\item \textit{Multi-lead time training:} We introduce a method for optimizing diffusion models for long lead times by training a single model to forecast multiple lead times (1h,  3h, and 6h). This improves long-range forecast skill by reducing cumulative error and encourages the model to learn a richer representation of atmospheric evolution across temporal scales. To our knowledge, this is the first application of multi-lead time training in the context of diffusion-based weather models. In contrast to StormCast, which is trained only for 1-hour horizons and rolled out autoregressively, our model is directly exposed to longer lead-time targets during training, improving robustness during extended inference. This simple yet effective strategy mitigates the compounding error problem that plagues single-step diffusion models, without requiring changes to the diffusion process or explicit noise scheduling. While recent work by \citet{cachay2025} addresses similar issues using complex architectures and lead-time-specific noise tuning, our approach demonstrates that long-range skill can also be achieved through lead-time-aware training.
\item \textit{True analysis data training:}  To enhance training quality, we use HRRR analysis fields as targets, rather than the 1-hour HRRR forecast fields used in StormCast.
\item \textit{GFS-guided downscaling:} Our model is conditioned and trained on Global Forecast System (GFS) forecasts as synoptic-scale input, rather than the ERA5 reanalysis dataset. While ERA5 is generally considered higher quality, \citet{adamov2025} found that models trained with ERA5 forcing can experience significant performance degradation when the forcing is replaced with forecasts from a global model such as IFS. They recommend training on ERA5 and fine-tuning on IFS. In contrast, our approach is to train the model entirely using GFS forcing, which also drives regional models like the Rapid Refresh (RAP) which in turn drives HRRR.
\item \textit{Future forcing data:} Unlike StormCast, which uses the GFS state at time $t$, our model incorporates the future GFS state (e.g., at time $t+1$ as input to the neural network). The GFS state at time $t$ may not provide much new information beyond what is already present in the HRRR analysis, though it could help align the model forecast with the global state during rollout\footnote{We thank Joel Oskarsson for useful discussion}. Prior work in \citet{larsson2025, adamov2025} has shown that incorporating both past and future boundary forcing improves model performance.
\end{itemize}

An important insight highlighted by \citet{adamov2025} is that models such as StormCast--and our own--can be viewed as hybrid models that combine forecasting with elements of downscaling. This is achieved by conditioning the model on full-domain GFS inputs, effectively treating the entire interior as a source of boundary information. Notably, this strategy is feasible only for machine learning-based Limited Area Models (LAMs); in contrast, traditional physics-based LAMs rely strictly on boundary conditions external to the domain for all times $t > 0$, though some implementations can impose additional constraints on large-scale features within the domain through nudging. However, the utility of this hybrid approach is not universal. For instance, the DANRA emulator examined by \citet{adamov2025} showed no benefit from including interior information, whereas their emulator for the COSMO model demonstrated substantial improvements. Furthermore, incorporating downscaling may degrade performance at shorter lead times. Therefore, a model that learns to dynamically blend forecasting and downscaling based on the forecast lead time may offer improved skill.

With these two components in mind, we initially considered enhancing the downscaling aspect by using a higher-resolution LAM that encompasses the HRRR domain -- specifically, the 12 km RAP model. In principle, statistical downscaling from the 12 km RAP to the 3 km HRRR alone should yield strong performance. However, we chose instead to use the 1/4 deg (28 km) resolution GFS data as conditioning, motivated by academic curiosity and the desire to avoid dependence on another LAM model -- particularly one like RAP, whose development has since been discontinued.

We explore two architectural variants in HRRRCast to evaluate how neural network design influences forecast skill at convection-allowing scales. Our primary model, ResHRRR, is a deep residual convolutional neural network enhanced with Squeeze-and-Excitation (SE) blocks and Feature-wise Linear Modulation (FiLM). It uses a Denoising Diffusion Implicit Model (DDIM) for probabilistic ensemble generation. Additionally, a single model is trained to predict multiple lead times to improve long-lead time forecasts. By integrating these design elements, ResHRRR aims to preserve fine-scale structure, reduce blurriness, and improve generalization across lead times. ResHRRR demonstrates improved skill over the operational HRRR forecast at the 20 dBZ composite reflectivity threshold, and competitive performance at 30 dBZ. As it forms the basis of most experiments and results presented in this work, we refer to this model as HRRRCast.

To understand how architectural choices influence forecast quality--and to assess the geometric flexibility of alternative designs--we also develop GraphHRRR, a Graph Neural Network (GNN) model inspired by GraphCast \citep{lam2023}. GraphCast has shown state-of-the-art performance in both global settings and limited-area domains, such as in WoFSCast \citep{Flora2025}. The pioneering work of \citet{keisler2022} employed a single-resolution graph, which was later extended in GraphCast to a multi-scale GNN that aggregates information across different spatial resolutions. However, the dense neighborhood connections in the multi-scale design can lead to spurious artifacts. To address this, \citet{oskarsson2023} proposed a hierarchical GNN over a rectangular grid to explicitly separate the processing of fine and coarse scales. While GraphHRRR is still under development, it maybe the future of HRRRCast due to its geometric flexibility \citep{adamov2025,oskarsson2023,keisler2022}. Looking ahead, we plan to integrate GNN-based encoders and decoders with fixed-grid processors such as ResNets or Transformers--as suggested by \citet{siddiqui2024}--to balance geometric adaptability with computational efficiency.

HRRRCast opens the door to augmenting HRRR with cost-effective, data-driven ensemble forecasts -- or potentially replacing it altogether once the system matures. Currently, the physics-based HRRR is computationally expensive to run, which has led to the adoption of unconventional ensemble techniques such as time-lagged ensembling, as used in the experimental HRRR-TLE \citep{xu2019}. Interestingly, a recent study by \citet{skinner2025} found that the primary advantage of the Warn-on-Forecast (WoFS) system -- a short-range regional model initialized with HRRR and run over a smaller domain -- lies in its probabilistic forecasts. HRRRCast has the potential to bridge this gap by enabling ensemble forecasts for HRRR; however, more research is needed to evaluate the effectiveness of such ensembling approaches.

\section{Methodology}

\subsection{Datasets}
Our training dataset comprises three years of HRRR analysis data and GFS data, with the latter used to provide synoptic-scale conditioning for the LAM. The training period spans March 2021 to March 2024, while the evaluation period covers four months, from April through July 2024. Additionally, a smaller evaluation dataset from May 1 to May 10, 2024, is used specifically for running computationally expensive diffusion models. This shorter period includes several major tornado outbreaks that occurred between May 6 and May 10, 2024 \citep{nws2024may7}.

Unlike \citet{pathak2024}, we chose to use analysis data because it is currently the highest-quality HRRR dataset available -- HRRR reanalysis data is not yet publicly accessible. From a theoretical standpoint, training on free-forecast data is likely to result in an emulator with lower skill than the HRRR model itself. However, we do not expect a dataset that is offset by +1 hour from the analysis time to perform significantly worse, since short-term forecasts are typically close to analyses. Infact, such forecasts may offer smoother and more temporally consistent fields than analyses, which can contain artifacts from data assimilation, potentially making them easier for neural networks to learn from. 

The variables used are similar to those in \citet{pathak2024}, with the main distinction being our use of GFS variables instead of the corresponding ERA5 variables (see Table \ref{inputs} for details). The HRRR dataset includes six atmospheric variables--temperature, geopotential height, horizontal and vertical wind components, and specific humidity--sampled at 12 pressure levels with higher resolution near the surface. Additionally, we include two key 2D variables, namely, 2-meter temperature and composite reflectivity (as a proxy for precipitation), and sea-level and surface pressures. For simplicity, we use pressure levels rather than HRRR's native hybrid levels, a choice also made by \citet{adamov2025} for one of their ML regional models (DANRA). In contrast, native sigma coordinates were used for their COSMO emulator.

The GFS dataset includes the same atmospheric variables, but sampled at four pressure levels, along with surface variables such as surface pressure, sea-level pressure, and composite reflectivity. Since our model performs both forecasting and downscaling, we find it beneficial to maintain consistency in variable types across HRRR and GFS inputs. However, due to an oversight, the GFS 2-meter temperature was not included during training.

\begin{table}
\centering
\caption{HRRRCast input variables and pressure levels.}
\begin{tabular}{p{6.2cm}p{4.5cm}p{4.5cm}}
\toprule
\textbf{Variables} & \textbf{HRRR levels (hPa)} & \textbf{GFS levels (hPa)} \\
\midrule
\textit{Atmospheric variables:} \newline
Temperature (\textbf{TMP}) \newline
Geopotential height (\textbf{HGT}) \newline
U component of wind (\textbf{UGRD}) \newline
V component of wind (\textbf{VGRD}) \newline
W component of wind (\textbf{WGRD}) \newline
Specific humidity (\textbf{SPFH}) &
200, 300, 475, 800, 825, 850, \newline
875, 900, 925, 950, 975, 1000 &
200, 500, 850, 1000 \\
\midrule
\multicolumn{3}{l}{\textit{Surface variables}} \\
Composite reflectivity (\textbf{REFC}) & Integrated & Integrated \\
2-meter temperature (\textbf{T2M} ) & Surface & \_ \\
Sea-level pressures (\textbf{MSLET}, \textbf{PRMSL}) &  \_ & Sea level \\
Surface pressure (\textbf{PRES}) & \_ & Surface \\
\midrule
\multicolumn{3}{l}{\textit{Static variables}} \\
Orography height (\textbf{OROG}) & Surface & \_ \\
Land mask (\textbf{LAND}) & Surface & \_ \\
\bottomrule
\end{tabular}
\label{inputs}
\end{table}

\subsection{Temporal and spatial resolution}
The HRRR is a high-resolution (3 km) model updated hourly, covering the CONUS domain with a grid size of 1059$\times$1799. Due to hardware limitations, we initially adopted the same configuration but on a smaller subdomain. Later, we opted for sampling every other gridpoint (i.e. 530$\times$900 grid with 6km resolution) to expand the domain to the entire CONUS. No explicit filtering was applied prior to subsampling -- our original intent was to preserve full resolution and rely on the neural network to learn appropriate scale representations. This allows the model to learn from storm activity across the U.S., which may outweigh the potential loss in accuracy from reduced resolution and the risk of aliasing fine-scale features.

The GFS runs every six hours at 00z, 06z, 12z, and 18z. Since the model requires HRRR and GFS inputs at each hourly timestep, we use GFS analysis data at synoptic times and GFS forecast data for the intervening hours. For simplicity, we regrid the GFS data to the HRRR grid and provide it to the neural network as additional input channels. Regridding could potentially be avoided by using a separate encoder head for the GFS input, mapping it into the same latent space as the HRRR encoder \citep{adamov2025}.

\subsection{Model architectures}
\subsubsection{ResHRRR architecture}
ResHRRR uses a deep residual neural network (ResNet) first introduced in \citet{he2015}. This architecture performs remarkably well in computer  vision tasks including weather prediction \citep{rasp2021}. ResNet addresses the problem of training very deep networks by using a novel architecture of which one of the key concept is residual learning. Each block in the ResNet learns the residual (difference between the input and output) via a skip connection. ResHRRR uses a variant named squeeze-and-excitation (SE)-ResNet, that significantly boosts the capabilities of ResNets models through the integration of channel-wise attention mechanisms. This can be important to capture non-linear relationships between different atmospheric and surface variables. For example, \citet{nguyen2023} found it beneficial to use channel-wise embedding even in Vision Transformers which were originally designed with the three RGB channels of images -- as opposed to the tens to hundreds of input channels encountered in weather maps. SE-ResNets dynamically recalibrate feature maps in each block of the ResNet by introducing a learnable fully connected layer that outputs scale factors, thereby pinpointing the most informative features to enhance both accuracy and efficiency. 

The FiLM block in ResHRRR is designed to incorporate conditioning information related to time, specifically the lead time and diffusion step, which are critical for DDPMs. This block processes the time-related embeddings through dense layers to generate scaling (gamma) and shifting (beta) parameters that modulate the intermediate feature maps within the network. ResHRRR applies the FiLM output across all major stages -- encoder, processor, and decoder blocks -- and the trainable skip connections described below. This widespread conditioning allows the network to effectively adapt its computations at each level of abstraction based on the diffusion stage and forecast lead time, which significantly enhances stability and accuracy during denoising, especially in later diffusion steps where precise temporal context becomes increasingly important. This design choice is inspired by findings in diffusion literature \citet{song2022, karras2022}, where strong time conditioning improves performance in stochastic generation tasks, and the Stormer global weather model architecture \citet{nguyen2023, nguyen2023b} where FiLM is used to improve training of a single model for forecasting multiple lead times.

Additional improvements include the introduction of two trainable skip connections that scale the HRRR input and its noised version before adding them to the output of the residual tower. This design offers two key benefits:
a) It enables the model to learn the difference from the current atmospheric state -- a strategy shown to be beneficial in several data-driven weather models.
b) It allows the model to adaptively weight the HRRR input and its noised counterpart as a function of the denoising step, inspired by \citet{karras2022}. The core idea is to give more weight to the original HRRR input during the early diffusion steps, and to the progressively denoised state in the later steps. This modification has proven effective in stabilizing training of the diffusion model.

\begin{figure}
\centering
\includegraphics[width=\linewidth]{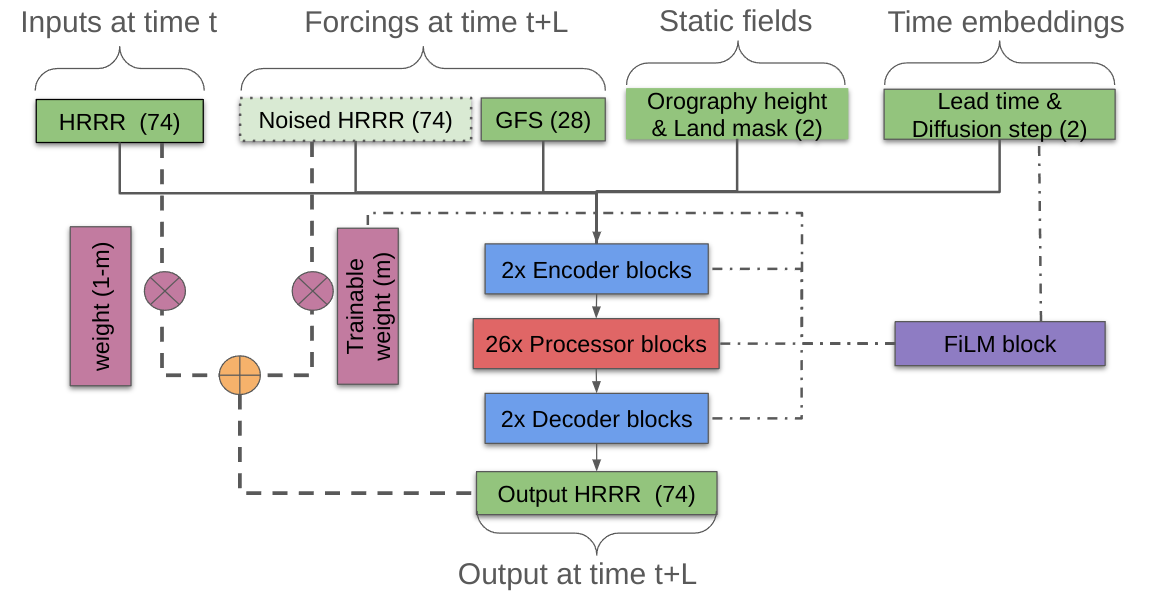}
\caption{ResHRRR inputs and architecture: The network accepts 180 channels (variables) with 530$\times$900 spatial dimensions and processes them through 30 SE-ResNet blocks. Additional improvements include FiLM block for time-conditioning and trainable skip connections for diffusion modelling.}
\label{res-arch}
\end{figure}

As mentioned earlier, it was not feasible to use the original HRRR domain dimensions of 1059$\times$1799. Instead, the network operates on input tensors of size 530$\times$900$\times$180. This input is quickly downsampled to a  latent space representation of size 130$\times$225$\times$192 using two U-net style encoder blocks -- each composed of SE-ResNet units with max pooling. All SE-ResNet blocks use layer normalization instead of batch normalization, which improves stability under small batch sizes and time-dependent conditioning. The latent representation is then processed by 26 SE-ResNet blocks (referred to as processor blocks), which retain the same spatial dimensions throughout. Finally, two decoder blocks upsample the output back to 530$\times$900$\times$74 using SE-ResNet units, with skip connections from the encoder concatenated at each stage, in line with the standard U-Net structure. The resulting model contains approximately 23.5 million trainable parameters.

\subsubsection{GraphHRRR architecture}
GraphHRRR is built on a GNN architecture originally developed for GraphCast \citep{lam2023}, and subsequently adapted for regional modeling in \citet{Flora2025} to create an emulator for the WoFS called WoFSCast. We refer readers to these works for a detailed description of the GNN architecture, and provide here a brief summary of the main architectural changes introduced in GraphHRRR compared to GraphCast: a) the icosahedral mesh refinement used for the global sphere is replaced with a Delaunay triangulation suitable for the rectangular domains of LAMs; b) periodic and cyclic boundary conditions along longitude/latitude boundaries are replaced with Dirichlet boundary conditions initialized using a global model; c) the latitude-longitude grid is replaced with HRRR's Lambert Conformal Conic (LCC) projection grid. The key advantage of this architecture lies in its geometric flexibility, making it the de facto choice for several global and regional data-driven weather models. It is especially well-suited for simulations on the sphere, offering improved accuracy over CNNs, and supports nested grid configurations with ease \citep{adamov2025,nipen2024}. Despite these strengths, GraphHRRR has several limitations -- primarily due to the complexity of modifying a sophisticated codebase like GraphCast. Currently, the model operates deterministically, does not ingest synoptic-scale inputs, and is not fine-tuned for long lead-time forecasts.

\subsubsection{Diffusion modeling}

The ResHRRR model leverages a DDPM framework, specifically the DDIM variant for faster ensemble generation. In this setup, the HRRR analysis at physical time $t$ is used as the conditioning input, while the model learns to generate HRRR at time $t{+}L$ through a learned reverse diffusion process. The forward process involves progressively adding Gaussian noise to the target HRRR state over $T=200$ steps, forming a trajectory from the data distribution to pure noise. During training, a diffusion step $\tau \in \{1, \ldots, T\}$ is randomly sampled for each training instance, and the model is trained to predict the clear image (equivalent to predicting the total noise ($\epsilon$) that was added).

The forward noising process is defined as:
\begin{equation}
x_\tau = \sqrt{\bar{\alpha}_\tau} \cdot x_0 + \sqrt{1 - \bar{\alpha}_\tau} \cdot \epsilon, \quad \epsilon \sim \mathcal{N}(0, I)
\end{equation}
where $\bar{\alpha}_\tau = \prod_{i=1}^\tau \alpha_i$ and $\alpha_i = 1 - \beta_i$. While $\beta_i$ is not used directly in the forward or reverse formulas, it underlies the construction of $\alpha_i$ in traditional DDPM formulations.

The $\beta$ values are scheduled using a cosine schedule:
\begin{equation}
\bar{\alpha}_\tau = \cos^2\left( \frac{\tau/T + s}{1 + s} \cdot \frac{\pi}{2} \right)
\end{equation}
where $T$ is the total number of diffusion steps (e.g., 200), and $s$ is a small offset (typically $s=0.008$) to prevent singularities at $\tau=0$. This yields smoother transitions compared to linear schedules, especially early in the diffusion process.

During inference, ResHRRR uses DDIM sampling, which is a non-stochastic, accelerated variant of DDPM. The update rule for one reverse step is:
\begin{equation}
x_{\tau-1} = \sqrt{\bar{\alpha}_{\tau-1}} \cdot \hat{x}_0 + \sqrt{1 - \bar{\alpha}_{\tau-1} - \sigma_\tau^2} \cdot \epsilon_\theta(x_\tau, \tau) + \sigma_\tau \cdot z
\end{equation}
where:
\begin{equation}
\hat{x}_0 = \frac{1}{\sqrt{\bar{\alpha}_\tau}} \cdot (x_\tau - \sqrt{1 - \bar{\alpha}_\tau} \cdot \epsilon_\theta(x_\tau, \tau)), \quad z \sim \mathcal{N}(0, I)
\end{equation}
In practice, we set $\sigma_\tau = 0$ for deterministic sampling or a small value for controlled stochasticity. In ResHRRR, inference typically runs for 30--50 steps, achieving high-fidelity forecasts efficiently.

While DDIM is computationally efficient and preserves sample quality in fewer steps, incorporating the Elucidated Diffusion Model (EDM) approach from Karras et al. \cite{karras2022} as used in \citet{pathak2024,price2024,chase2025} could offer advantages such as dynamic noise scaling, optimal time step discretization, and improved sample diversity and stability---especially under high noise levels. EDM's use of $\sigma$-based noise levels rather than $\beta$-based scheduling may allow for more fine-grained control over generation difficulty and quality, which could be beneficial in the chaotic and data-rich context of weather modeling.

\subsection{Training}

\subsubsection{Training objective}
The training objective is to minimize the mean squared error (MSE) between the network predictions and the ground truth, which is the HRRR analysis data. In both ResHRRR and GraphHRRR, we adopt the pressure-weighted loss strategy introduced by \citet{lam2023}, where greater weight is assigned to variables near the surface. For each of the six atmospheric variables, the weights across pressure levels sum to 1. Additionally, the 2-meter temperature and composite reflectivity are assigned weights of 0.1 and 1.0, respectively. Latitude weighting is also applied, however, its effect in LAM domains away from the poles is minimal.

\subsubsection{Multi-lead time training}
There are two main approaches to improving long-lead time forecasts: (a) training a model to predict multiple lead times \citep{bi2023,nguyen2023}, and (b) fine-tuning a model trained on a single lead time by accumulating loss over rollout trajectories \citep{lam2023}. However, rollout-based fine-tuning strategies do not translate naturally to diffusion-based modeling as evidenced by the lack of fine-tuning in several probabilistic models including \citet{pathak2024,price2024}. It is worth noting that global data-driven models, which lack additional conditioning input, tend to rely more heavily on fine-tuning to maintain skill at longer lead times. In contrast, LAMs that incorporate full-domain synoptic data as conditioning input benefit from a more grounded forecast state, which helps mitigate the accumulation of errors during auto-regressive rollouts to some extent.

ResHRRR trains a single model to predict multiple lead times (1h, 3h, and 6h). Instead of training separate models for each lead time, the model is trained with batches uniformly sampled over these three lead times, each given equal weight, which encourages the network to learn representations that generalize across lead times. Although predicting the 6h lead involves larger errors and thus tends to dominate the loss function during training, this bias aligns well with practical rollout scenarios where longer lead time predictions are repeatedly used -- for example, generating a 38-hour forecast requires rolling out six 6h steps followed by a final shorter step. This iterative rollout strategy is similar to that employed by the Pangu \citep{bi2023} and Stormer \citep{nguyen2023} models, leveraging iterative dynamics forecasting to extend forecasts beyond the training horizon. Additionally, ResHRRR incorporates the lead time information explicitly into the model through FiLM blocks, strengthening the model’s ability to adapt its predictions depending on the forecast lead time. This multi-lead training framework helps mitigate the degradation in accuracy typically observed when a model trained only for 1h lead times is applied at longer horizons.

While recent work such as Elucidated Rolling Diffusion Models (ERDM) \citep{cachay2025} targets long-lead forecast skill by predicting full trajectories in a single forward pass using time-aware noise schedules and lead-time-specific loss weighting, the Rolling Diffusion framework \citep{ruhe2024} applies similar ideas in the context of DDPMs, which is also the formulation we use in HRRRCast. Rolling Diffusion introduces a sliding-window denoising approach where future frames are assigned increasing noise levels based on temporal distance from conditioning input. ERDM extends this concept to Elucidated Diffusion Models (EDMs), combining it with trajectory-aware conditioning and stochastic sampling. While these approaches elegantly address long-range uncertainty accumulation, they introduce added architectural and computational complexity. In contrast, our multi-lead-time training strategy improves long-horizon performance in a simpler and model-agnostic way. By jointly training on 1h, 3h, and 6h lead times and explicitly conditioning on lead time via FiLM blocks, ResHRRR learns temporal generalization directly, without requiring modifications to the diffusion process or reliance on external forecasts.

\subsection{Ensemble generation and mean computation}
In traditional physics-based regional models like the HRRR, ensemble members are generated using a variety of perturbation techniques to capture forecast uncertainty. These include perturbing initial conditions and applying stochastic physics schemes such as Stochastically Perturbed Parameterization Tendencies (SPPT) and Stochastic Kinetic Energy Backscatter (SKEB), which account for model uncertainty by randomly modifying physical tendencies or injecting energy at unresolved scales \citep{jankov2019}. Analogous strategies are emerging in diffusion-based, data-driven models. For instance, using different initial conditions is directly applicable by perturbing the input fields or conditioning variables. The addition of Gaussian noise at the start of the reverse diffusion process resembles the Ensemble of Data Assimilation (EDA), while injecting noise during each denoising step serves as a proxy for modeling internal uncertainty, much like SPPT or SKEB in physics-based systems. For the results presented in this work, ensemble members are generated by perturbing only the initial Gaussian noise since we found that to produce adequate forecast spread (see \ref{append_spread}). We evaluate from 3 to 10 ensemble members generated using this process. While deterministic data-driven models are generally very fast to run, probabilistic forecasting using diffusion models introduces 20–50 sampling steps per forecast, making a 10-member ensemble at least two orders of magnitude slower.

Interpreting ensemble forecasts can be challenging, especially when dealing with large ensembles. Communicating the forecast and associated uncertainty to end users becomes difficult due to the volume and complexity of the information. To address this, a common strategy is to compute a single deterministic forecast--typically an ensemble mean--that synthesizes information across all members. Several averaging methods exist for precipitation-related variables like composite reflectivity. A simple arithmetic mean often yields forecasts with low RMSE but tends to filter out small-scale features due to lack of member consensus, making it less suitable for precipitation forecasts where such details matter. Alternatively, a maximum average approach captures peak values more aggressively but tends to overestimate intensities, introducing high bias. The Probability Matched Mean (PMM) has emerged as a preferred method for reflectivity, as it reconstructs the intensity distribution by matching the rank-ordered values from the ensemble, preserving both spatial structure and amplitude \citep{clark2017}.

\section{Results and discussion}
To evaluate the skill of HRRRCast, we use two different datasets: a) The first is a small test dataset spanning May 1 to May 10, 2024, which includes several convective events. Forecasts on this dataset are generated at hourly intervals, similar to the operational HRRR, resulting in a total of 240 timestamps. Each forecast is run up to 18 hours lead time. b) The second dataset covers the full 4-month testing period, with forecasts initialized every 6 hours. This results in approximately 488 timestamps. Since HRRR runs forecasts up to 48 hours at 00Z, 06Z, 12Z, and 18Z, we follow the same schedule for this dataset as well.

The HRRR analysis data, which served as the training target, is used as the ground truth for evaluation. While the Multi-Radar Multi-Sensor (MRMS) dataset provides a more accurate source of composite reflectivity, we did not observe significant differences between it and the HRRR analysis. Using MRMS as ground truth would have been more appropriate if the model had been trained directly on MRMS, or alternatively, if it had been trained on HRRR forecast data, as done in \citet{pathak2024}.

We begin with qualitative comparisons of forecasts at selected initialization times, followed by a quantitative evaluation focused primarily on the Fractions Skill Score (FSS) -- a neighborhood-based metric -- for reflectivity. In addition, we compute standard grid-based metrics derived from the 2$\times$2 contingency matrix, using the same thresholds applied in the FSS calculation. Finally we present preliminary results with object-based verification.

Variables other than composite reflectivity are evaluated using the root-mean-square error (RMSE) metric.

\subsection{Qualitative evaluation}
\begin{figure}
\centering

\noindent
\makebox[0.22\linewidth][c]{HRRR analysis}%
\hfill
\makebox[0.22\linewidth][c]{HRRR forecast}%
\hfill
\makebox[0.22\linewidth][c]{HRRRCast 5-mem.}%
\hfill
\makebox[0.24\linewidth][c]{HRRRCast 1-mem.}%

\vspace{0.5em}

\setlength{\tabcolsep}{4pt} 
\begin{tabular}{@{} l c @{}}
  \includegraphics[width=\linewidth]{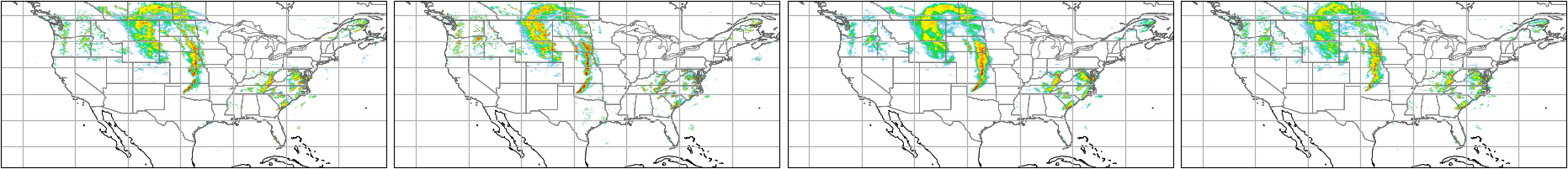} & f01 \\[6pt]
  \includegraphics[width=\linewidth]{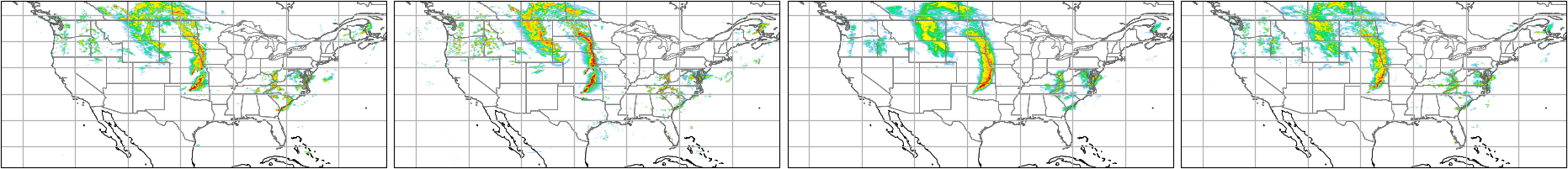} & f03 \\[6pt]
  \includegraphics[width=\linewidth]{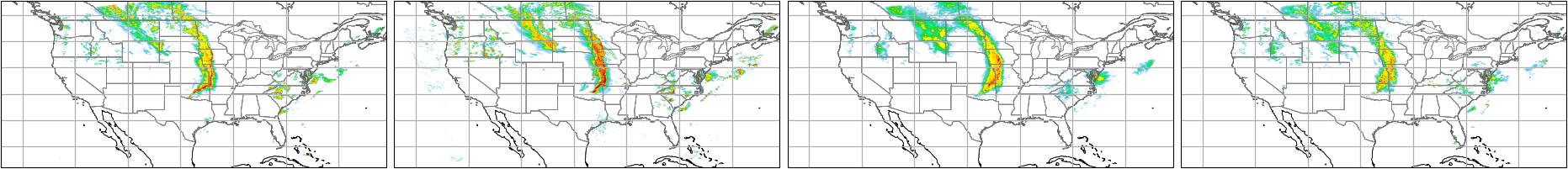} & f06 \\[6pt]
  \includegraphics[width=\linewidth]{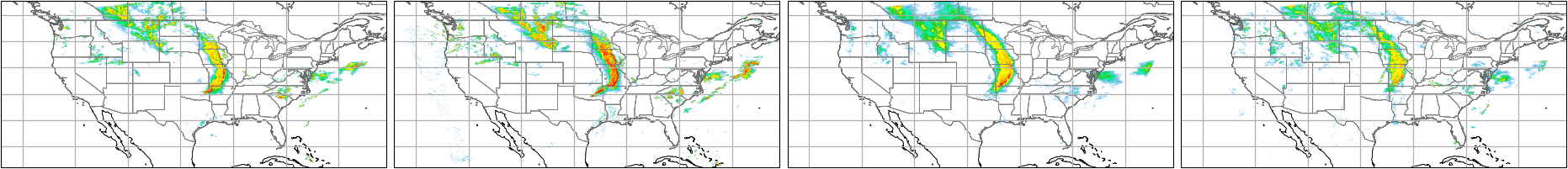} & f09 \\[6pt]
  \includegraphics[width=\linewidth]{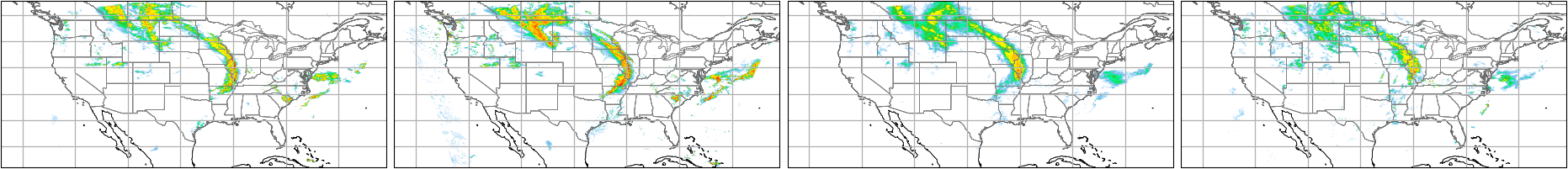} & f12 \\[6pt]
  \includegraphics[width=\linewidth]{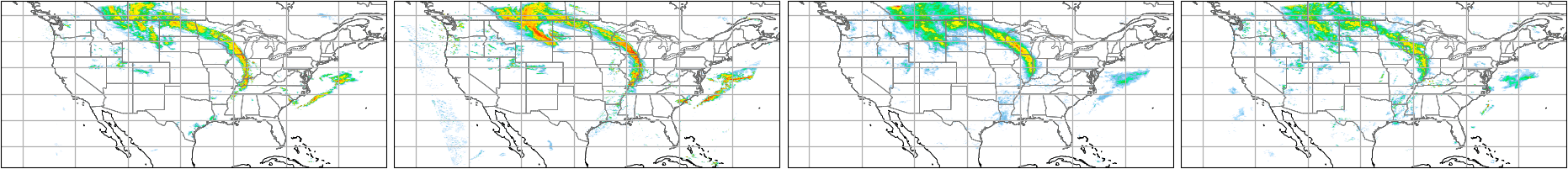} & f15 \\[6pt]
  \includegraphics[width=\linewidth]{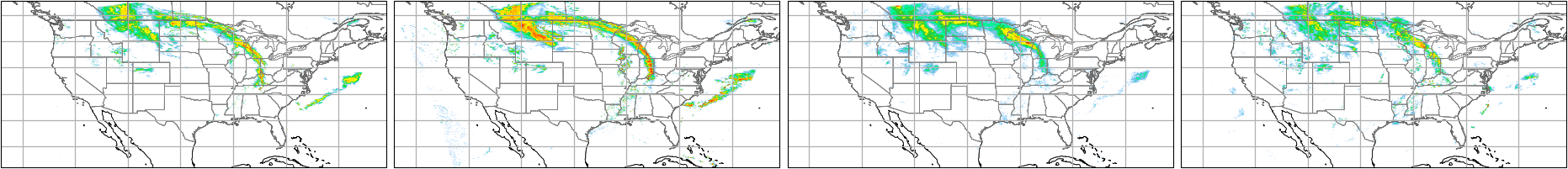} & f18 \\ [6pt]
\multicolumn{2}{c}{
  \begin{minipage}{0.4\linewidth}
    \centering
    \includegraphics[width=\linewidth]{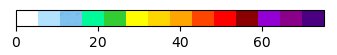} \\
    \small Composite reflectivity (dBZ)
  \end{minipage}
} \\
\end{tabular}

\caption{Composite reflectivity at forecast initialization time 2024-05-06 23:00 UTC. Shown are the HRRRCast single-member forecast and the 5-member ensemble Probability-Matched Mean (PMM), compared against the HRRR forecast and the ground truth--the HRRR analysis.}
\label{refc_map}

\end{figure}

Figure \ref{refc_map} presents a qualitative comparison of HRRRCast ensemble forecasts (up to 5 members), the HRRR model forecast, and the ground truth, for lead times up to 18 hours. HRRRCast produces visually realistic forecasts across all lead times. Notably, the blurriness that often affects deterministic data-driven weather models is significantly reduced in HRRRCast, owing to its diffusion-based generative architecture. However, some smoothing is still apparent at longer forecast times.

In contrast, the HRRR model tends to overestimate composite reflectivity, a trend evident in both Figure \ref{refc_map} and the additional comparisons shown in Figure \ref{refc_map2} and \ref{refc_map3}. HRRRCast forecasts, however, align more closely with the HRRR analysis, indicating improved calibration and more realistic spatial structure in the predicted reflectivity fields. That said, \citet{Flora2025} observed that training their model on forecast data improved reflectivity sharpness for lead times up to 2 hours. While the analysis dataset is of higher quality--especially relevant for HRRRCast, which supports forecast lead times up to 48 hours--it remains an open question whether incorporating a mix of forecast and analysis data could enhance performance, particularly at the 40 dBZ reflectivity threshold. We plan to explore this in future work.

\begin{figure}
\centering

\noindent
\makebox[0.22\linewidth][c]{HRRR analysis}%
\hfill
\makebox[0.22\linewidth][c]{HRRRCast 10-mem.}%
\hfill
\makebox[0.22\linewidth][c]{ResHRRR determ.}%
\hfill
\makebox[0.24\linewidth][c]{GraphHRRR determ.}%

\vspace{0.5em}

\setlength{\tabcolsep}{4pt} 
\begin{tabular}{@{} l c @{}}
  \includegraphics[width=\linewidth]{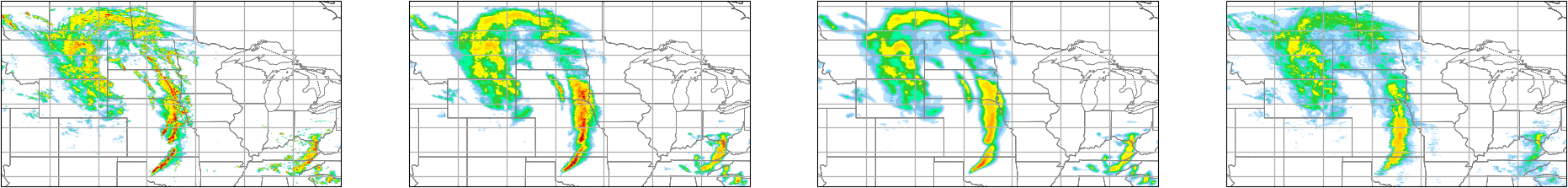} & f01 \\[6pt]
  \includegraphics[width=\linewidth]{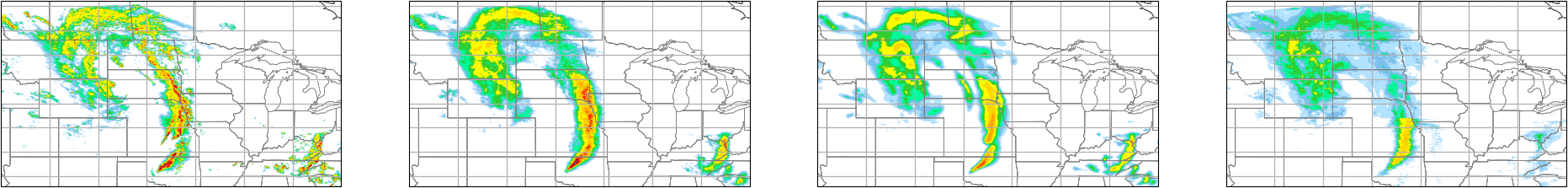} & f02 \\[6pt]
  \includegraphics[width=\linewidth]{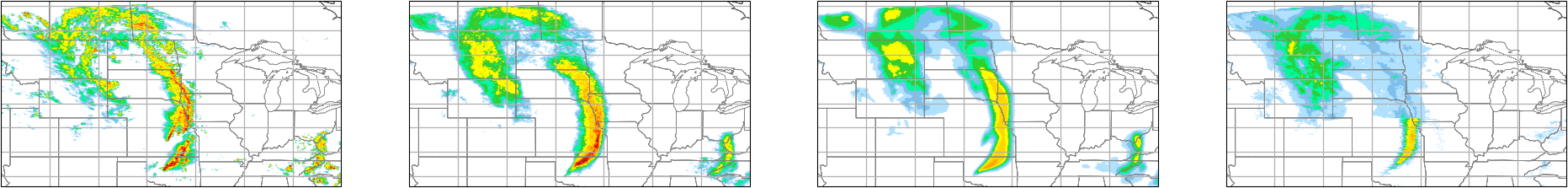} & f03 \\[6pt]
  \includegraphics[width=\linewidth]{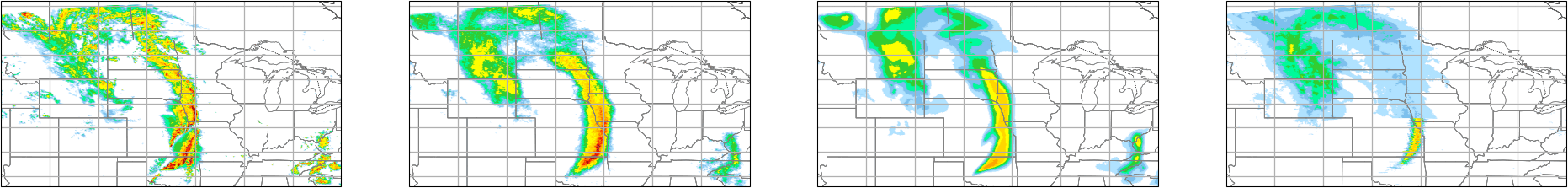} & f04 \\[6pt]
  \includegraphics[width=\linewidth]{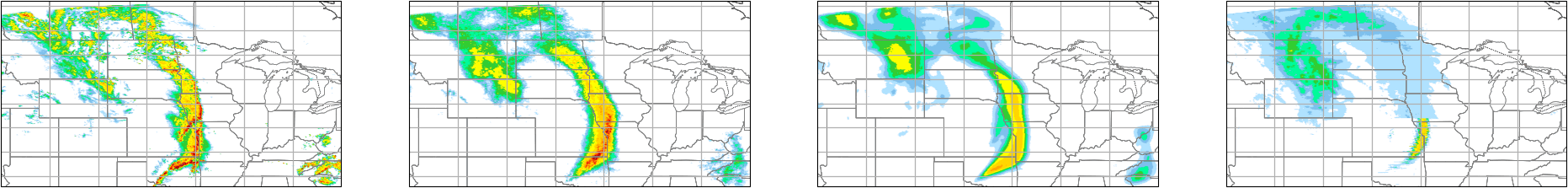} & f05 \\[6pt]
  \includegraphics[width=\linewidth]{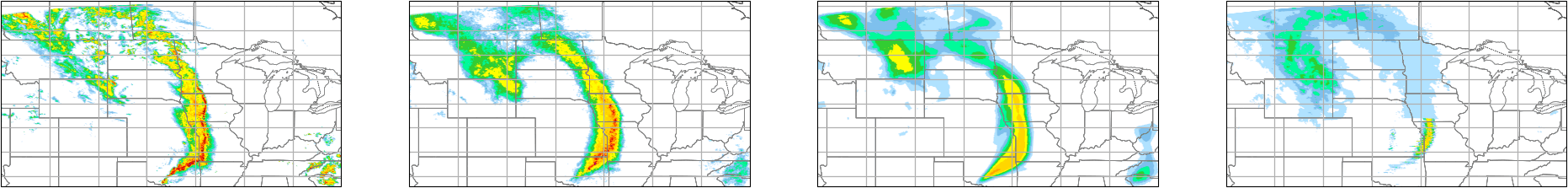} & f06 \\[6pt]
  \includegraphics[width=\linewidth]{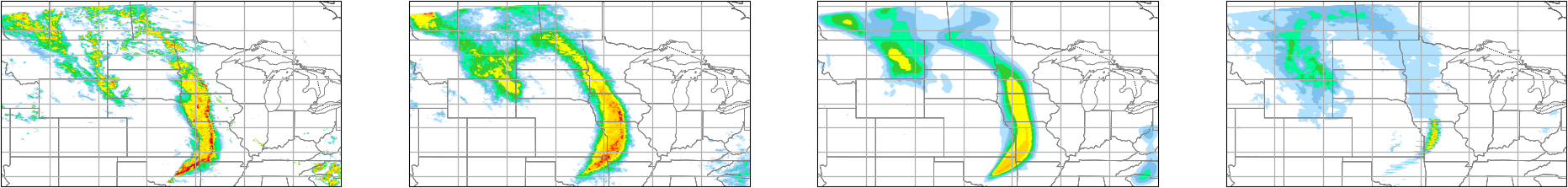} & f09 \\[6pt]
\multicolumn{2}{c}{
  \begin{minipage}{0.4\linewidth}
    \centering
    \includegraphics[width=\linewidth]{pics/refc-colorbar.png} \\
    \small Composite reflectivity (dBZ)
  \end{minipage}
} \\
\end{tabular}

\caption{Enlarged view of composite reflectivity plots for 10-member HRRRCast ensemble and deterministic models at forecast initialization time of 2024-05-06 23:00 UTC. The blurriness typical of deterministic forecasts is clearly visible. In particular, the GraphHRRR model's lack of fine-tuning and/or multi-lead-time training has negatively affected its performance at modest forecast lead times of 6 and 9 hours.}
\label{refc_map_det}

\end{figure}

Figure \ref{refc_map_det} presents results from the deterministic versions of our ResNet-based and graph-based models. Both deterministic models exhibit noticeable blurring, even at relatively short lead times such as 6 hours. This issue is particularly pronounced in the GraphHRRR model, which begins to blur rapidly due to having been trained only for 1-hour lead times without fine-tuning,  and then requiring six rollouts to generate a 6-hour forecast, compounding the degradation in forecast quality.

\subsection{Fractions Skill Score}
\begin{figure}[htbp]
\centering
\begin{subfigure}[t]{\textwidth}
    \centering
    \includegraphics[width=0.327\linewidth]{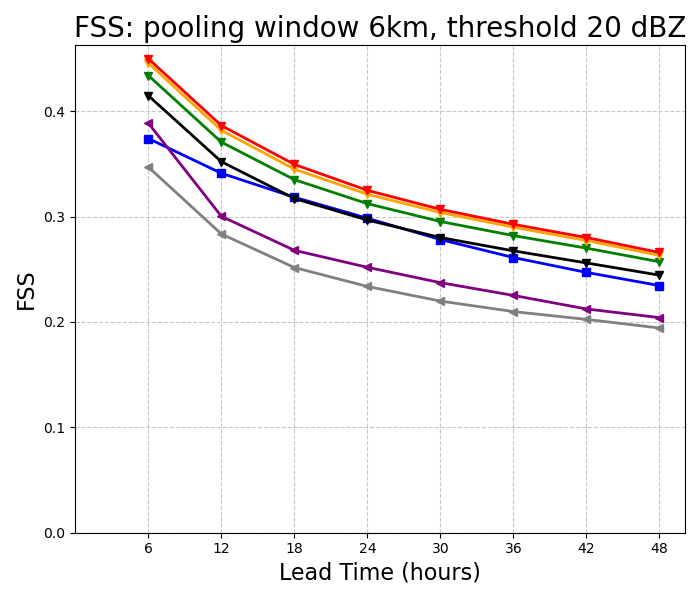}
    \includegraphics[width=0.327\linewidth]{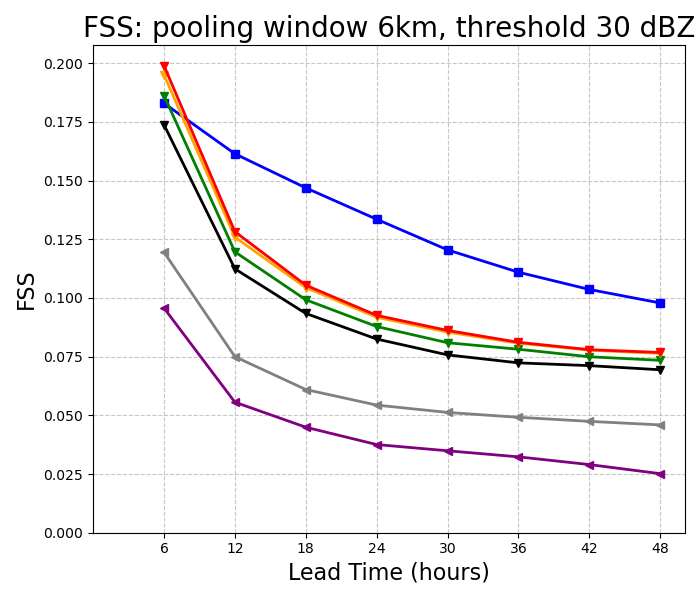}
    \includegraphics[width=0.327\linewidth]{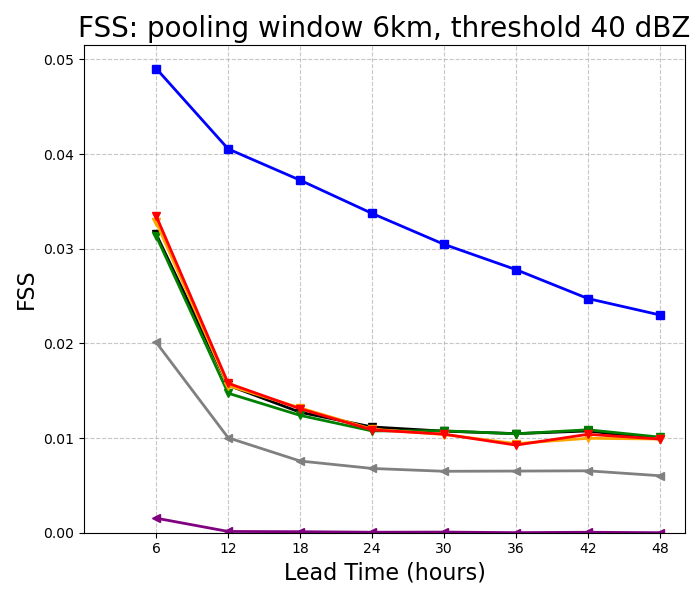}
    \caption{4-month dataset at 6-hourly intervals and forecasts upto 48 hrs.}
    \label{fss-4month}
\end{subfigure}
\hfill
\begin{subfigure}[t]{\textwidth}
    \centering
    \includegraphics[width=0.327\linewidth]{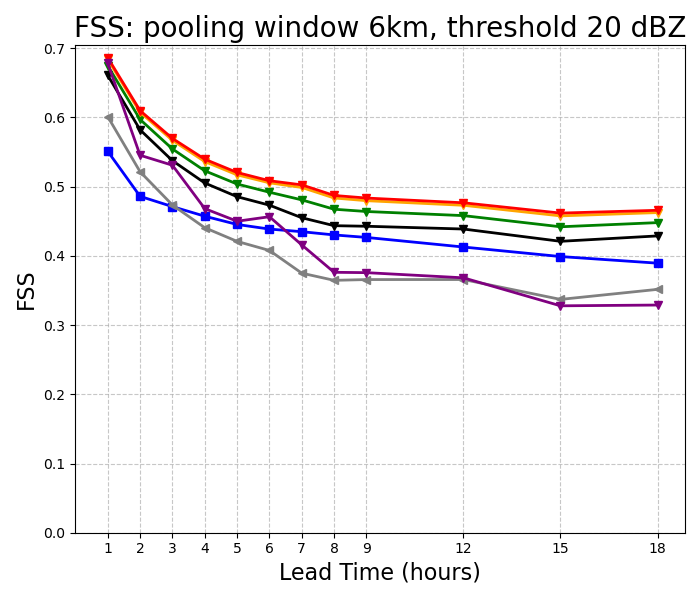}
    \includegraphics[width=0.327\linewidth]{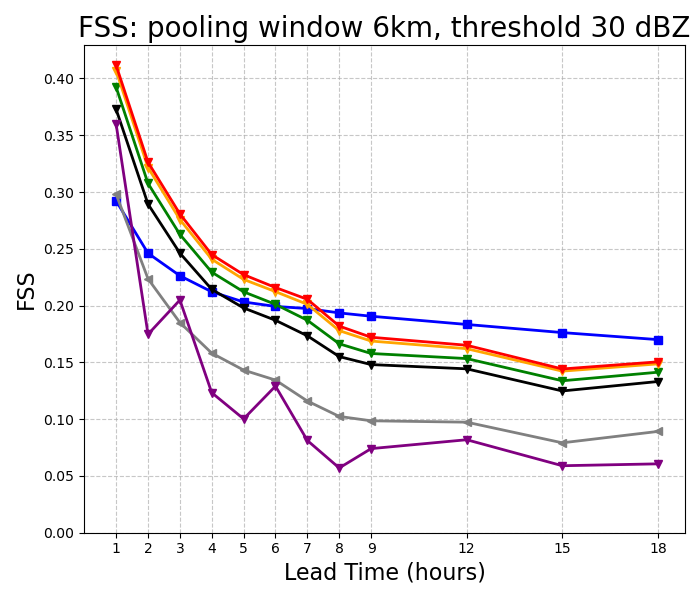}
    \includegraphics[width=0.327\linewidth]{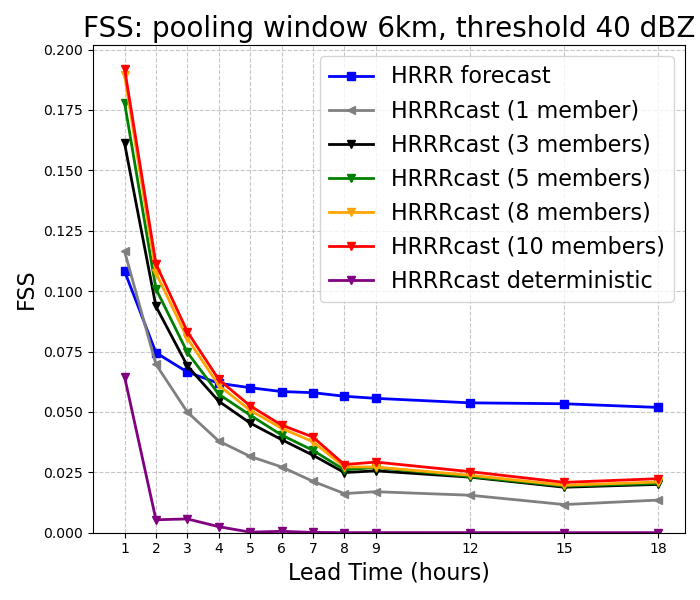}
    \caption{10-day dataset at hourly intervals and forecasts upto 18 hrs}
    \label{fss}
\end{subfigure}
\hfill
\begin{subfigure}[t]{\textwidth}
\centering
\includegraphics[width=0.327\linewidth]{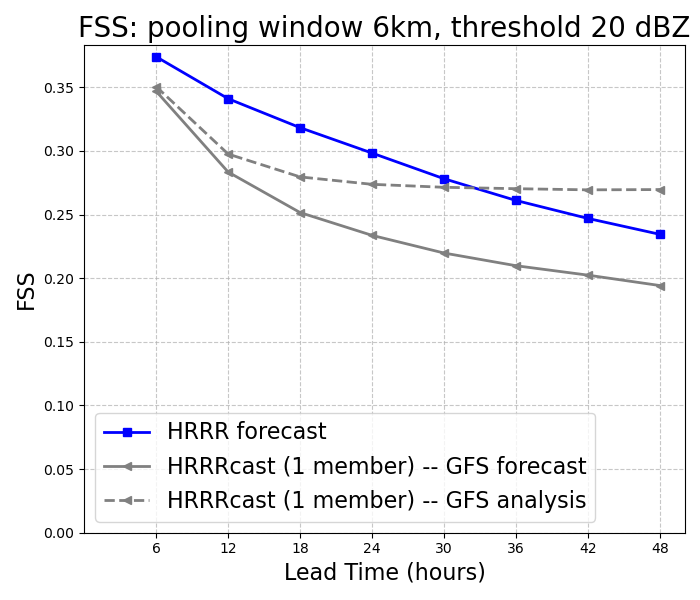}
\includegraphics[width=0.327\linewidth]{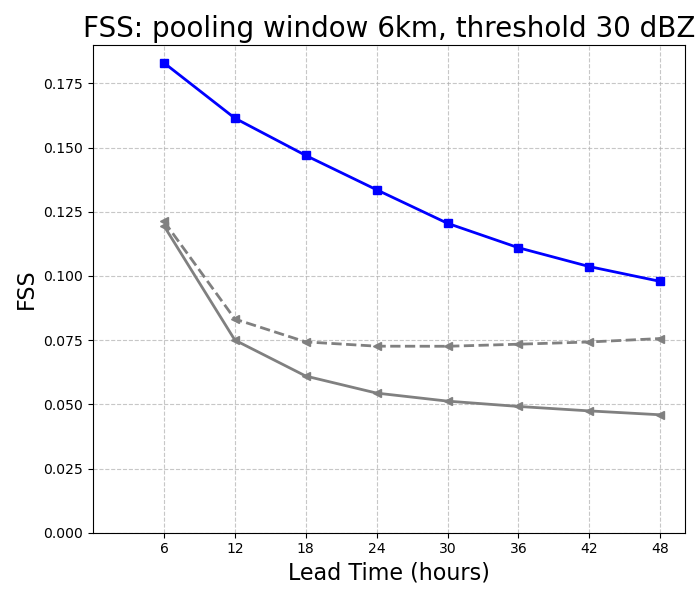}
\includegraphics[width=0.327\linewidth]{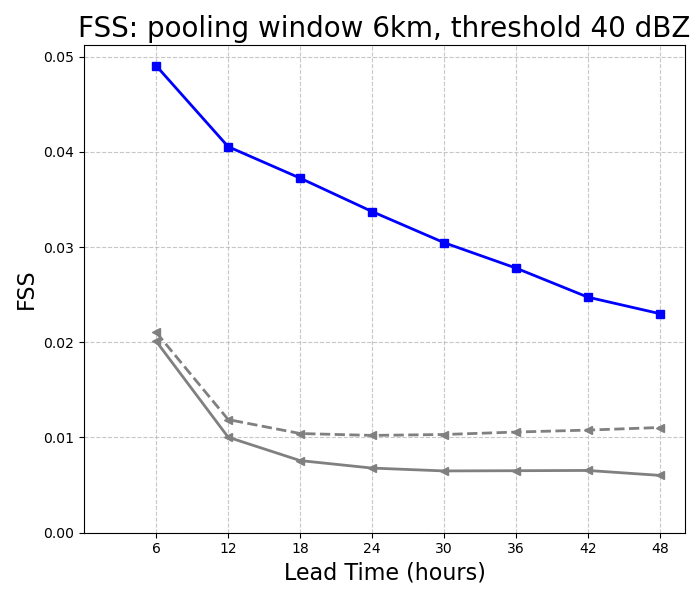}
\caption{Comparing GFS forecast data vs GFS analysis data for synoptic conditioning.}
\label{fss-analysis}
\end{subfigure}
\caption{Skill scores of composite reflectivity with different datasets, different thresholds (20, 30, and 40 dBZ) and  a 6 km pooling window (grid scale)}
\label{fig:fss-combined}
\end{figure}

Next, we quantitatively evaluate the performance of HRRRCast using the Fractions Skill Score (FSS) metric computed with 6km pooling window (the grid scale). Figure \ref{fss} and \ref{fss-4month} show that HRRRCast performs very well at light rainfall (20 dBZ threshold), outperforming HRRR at all lead times up to 48 hours, even when using only three ensemble members. This represents an improvement over the state-of-the-art StormCast model \citep{pathak2024}, which was generally able to match HRRR only at lead times of 6 hours or less.

For Figure \ref{fss-4month}, we do not show results for the first 6 hours due to the 6-hourly forecast initialization interval. However, HRRRCast is generally competitive with, or outperforms, HRRR within this range, even in its least performant configuration that excludes synoptic-scale input.

The forecast skill of HRRRCast declines rapidly beyond 6 hours when GFS input data is omitted, underscoring the importance of incorporating synoptic-scale information to improve performance at longer lead times. While we have not yet conducted a formal ablation study to confirm this, the inclusion of composite reflectivity among the GFS-derived inputs (absent in StormCast) may also contribute to the enhanced performance. We hypothesize that using a consistent set of variables in both the synoptic input (GFS) and the mesoscale input (HRRR) improves the model’s downscaling capability. This hypothesis is supported by the design of alternative approaches to limited-area modeling -- such as nested-grid and stretched-grid simulations \citep{nipen2024} -- which also rely on consistent variables across both the global and refined domains.

HRRRCast also performs well at the light-to-moderate rainfall threshold (30 dBZ), outperforming HRRR up to 7-hour lead times, aided by the use of ensembles. However, beyond this range, HRRR regains an advantage.

It is important to note that all models exhibit ``usable skill" (defined as FSS $\ge$ 0.4) only at larger pooling windows and relatively short lead times (see Figures \ref{fss-others} and \ref{fss-others-4month}). HRRR tends to perform better at larger pooling windows, whereas HRRRCast shows superior performance at the grid scale. This suggests that HRRRCast provides better spatial precision and storm placement accuracy than HRRR. The FSS metric becomes increasingly tolerant of displacement errors as the pooling radius increases.

At the moderate rainfall threshold (40 dBZ), none of the models achieve usable skill. In this regime, HRRR performs slightly better than HRRRCast, likely due to its tendency to over-predict precipitation, as discussed earlier. \citet{Flora2025} similarly hypothesize that training AI models on forecast data--rather than on analysis--can produce sharper and more intense predictions. This may be because storm-scale analyses often contain errors, such as missing or misrepresented storms, which complicate the learning process. Additionally, reducing temporal resolution through subsampling has been shown to overly dissipate small-scale features \citep{smith2023}, further limiting model performance at higher thresholds. These factors may help explain why HRRR and forecast-trained AI models tend to perform relatively better in this regime. Future improvements could include training on the 15-minute resolution HRRR dataset and exploring a hybrid training approach that mixes HRRR analyses with forecasts.

The deterministic version of HRRRCast performs very well on the RMSE metric -- achieving even lower RMSE than the 10-member ensemble of the probabilistic HRRRCast on most variables (see Figure \ref{rmse_comp}) -- but it lags behind in terms of the Fractions Skill Score (FSS) for composite reflectivity, particularly at higher dBZ thresholds where blurriness has the most impact. This blurriness is especially detrimental for variables with sharp spatial gradients and localized extremes, such as precipitation and vertical velocity, while the deterministic model performs well on smoother fields like temperature and geopotential height.

To address these limitations, \citet{subich2025} introduced a modified loss function that preserves fine-scale features and mitigates the “double penalty” problem in deterministic forecasts. Another promising direction is the use of CRPS-based training, as in AIFS-CRPS \citep{lang2024}, which enables sharper and more realistic ensemble members by optimizing a probabilistic loss rather than MSE. Incorporating such loss functions and architectural improvements into future versions of deterministic HRRRCast may help bridge the gap with diffusion-based probabilistic models.

Another question we aim to address is the impact of synoptic data quality on forecast performance at longer lead times. In Figure \ref{fss-analysis}, we compare the use of GFS forecast data with GFS analysis data (the best available). The results suggest three distinct regimes: a) For lead times less than 6 hours, the model benefits more from forecasting directly from the high-resolution initial conditions than from downscaling. In this regime, synoptic-scale information offers little benefit and may even negatively impact performance. b) Between 6 and 18 hours, the influence of synoptic-scale data begins to emerge. However, the performance difference between using GFS forecast data versus GFS analysis data remains relatively modest. c) Beyond 18 hours, the advantage of using analysis data becomes more pronounced. In this range, performance when using analysis data remains relatively stable, indicating that the model is effectively downscaling from the high-quality analysis fields. These findings are consistent with those of \citet{adamov2025}, who observed that the skill obtained from downscaling a high-quality dataset like ERA5 -- as opposed to a free-running forecast such as IFS -- should persist indefinitely due to the incorporation of future observations. While using analyses for operational forecasting is not practical, this result underscores the potential benefit of improved global models (e.g., data-driven models) for enhancing long lead-time forecasts.

\subsection{Metrics based on contingency table}
Figures \ref{metrics_anal}  shows results of standard metrics such as Probability of Detection (POD), Success Ratio (SR), Critical Success Index (CSI), Frequency Bias (FB) computed using grid based verification approach. Here again, HRRRCast ensemble outperforms HRRR forecast on the CSI metric at 20 dBZ and 30 dBZ thresholds. 

The performance comparison between HRRR and HRRRCast  reveals distinct characteristics in storm prediction behavior across lead times. Most notably, the FB plot  shows that HRRR systematically overpredicts storm occurrences, with FB values ranging from approximately 1.2 to over 2.7 depending on the reflectivity threshold. This indicates that HRRR forecasts significantly more storm events than are observed, especially at lower thresholds (e.g., 20 dBZ). In contrast, HRRRCast consistently underpredicts, with FB values well below 1 (typically around 0.4 to 0.8), showing a tendency to miss events. The green dashed line at FB = 1.0 marks perfect frequency bias, which neither system achieves, though HRRRCast is closer to unbiased in some mid-range thresholds.Additionally, Figure \ref{metrics_mrms} shows performance metrics computed using MRMS as the ground truth. The plot confirms that HRRRCast more closely matches observations than HRRR in terms of frequency bias.

In the Performance Diagram, HRRR achieves moderately higher POD values compared to HRRRCast, but at the cost of a lower SR, reflecting its overforecasting nature (i.e., high false alarm rate). The points for HRRR lie above the diagonal FB = 1.0 line, confirming the overprediction behavior. Conversely, HRRRCast (especially the 10-member ensemble in red) shows better balance between POD and SR, with points closer to the diagonal, but lower overall POD -- it is more conservative and less noisy, but misses more events. In short, HRRR is aggressive and overpredicts, while HRRRCast is conservative and underpredicts, and the choice between them depends on whether missed events or false alarms are more operationally costly.

\subsection{Object-based verification}
As a complement to the grid-based and neighborhood-based verification already performed, we conducted preliminary object-based verification of the 10-member HRRRCast ensemble, comparing it against both the deterministic HRRR forecast and the HRRR analysis. Using the MODE (Method for Object-based Diagnostic Evaluation) tool within the MET verification framework\citep{metplus2024}, we analyzed attributes such as object centroid location, intensity distribution, area, and orientation. Initial comparisons indicate that HRRRCast effectively captures the spatial structure and variability of convective systems, performing at least as well as the HRRR model.

\begin{figure}
\centering
\begin{subfigure}[t]{\textwidth}
\centering
\includegraphics[width=\linewidth]{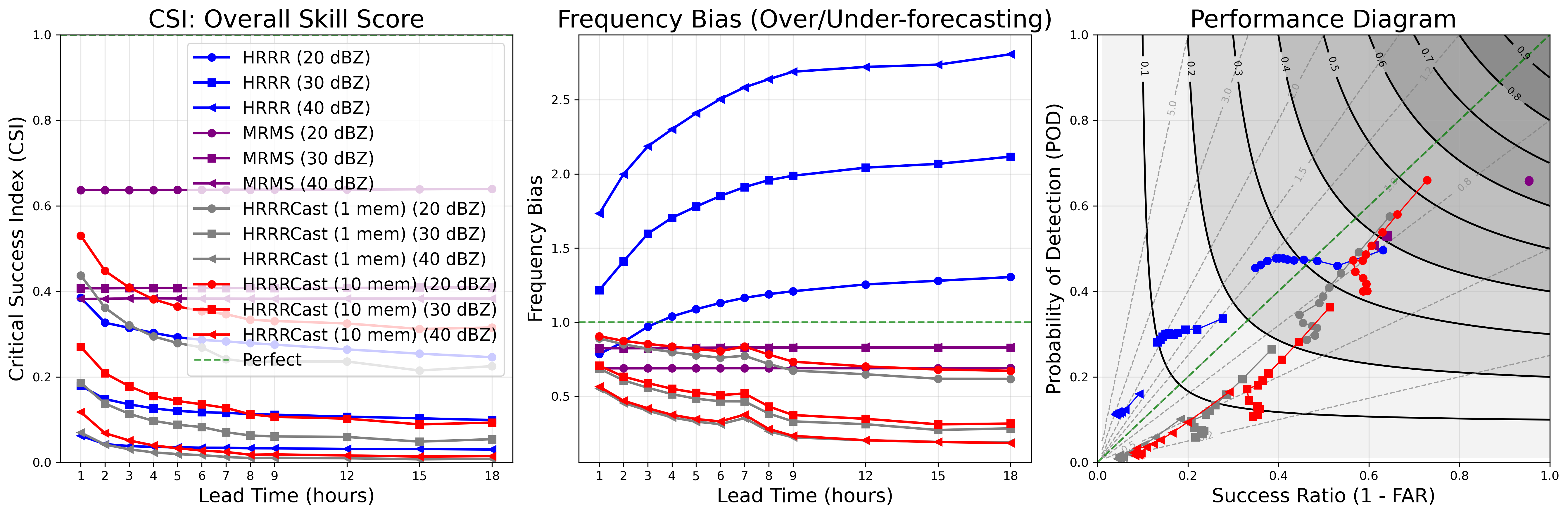}
\caption{Grid-based verification using HRRR analysis as ground truth}
\label{metrics_anal}
\end{subfigure}
\par\vspace{0.5cm}\noindent
\begin{subfigure}[t]{\textwidth}
\centering
\includegraphics[width=\linewidth]{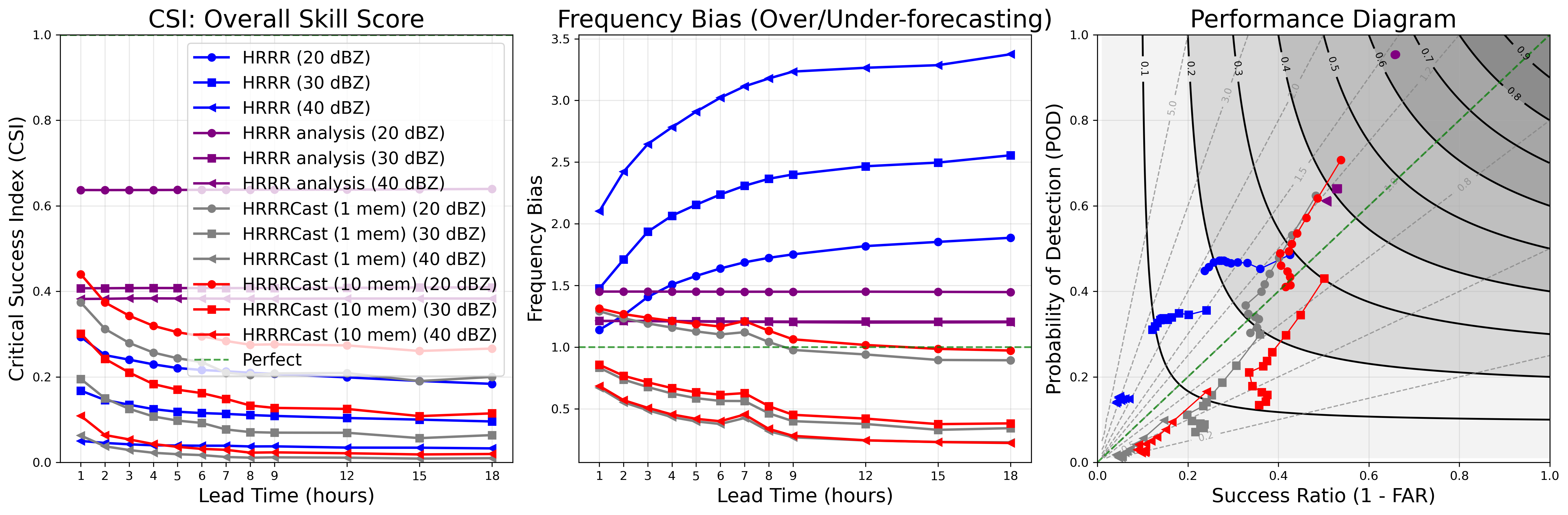}
\caption{Grid-based verification using MRMS as ground truth}
\label{metrics_mrms}
\end{subfigure}
\par\vspace{0.5cm}\noindent
\begin{subfigure}[t]{\textwidth}
\centering
\includegraphics[width=\linewidth]{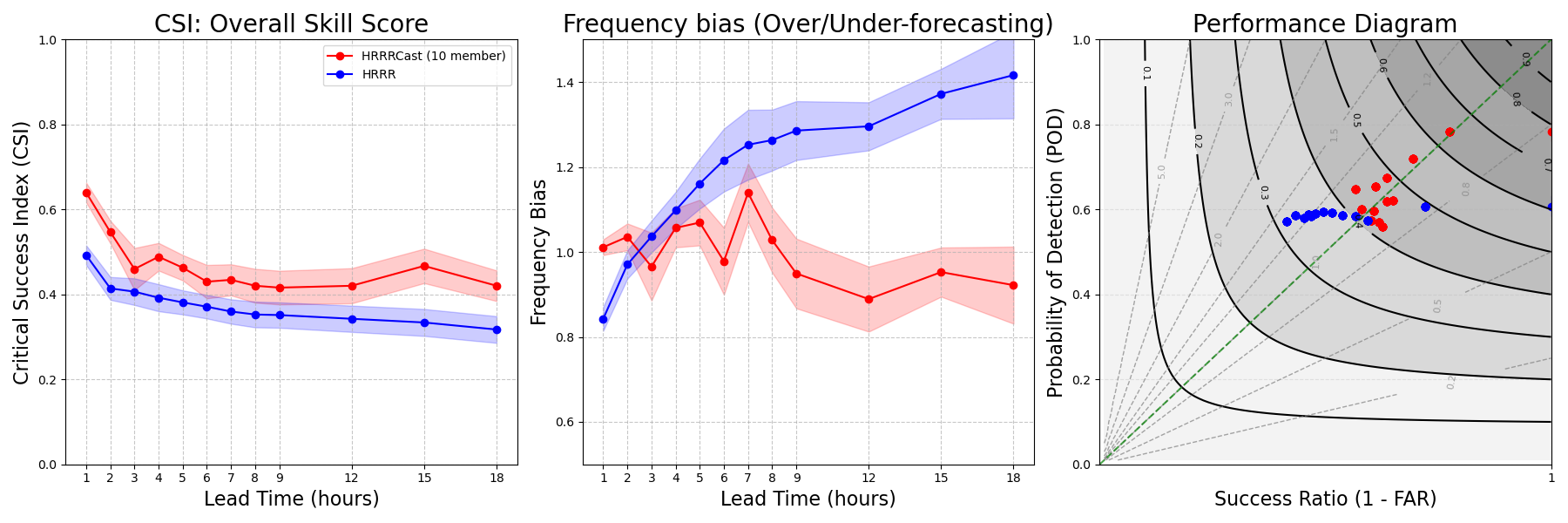}
\caption{Preliminary object-based verification using HRRR analysis as ground truth.}
\label{obj-metrics}
\end{subfigure}
\caption{Standard metrics comparing HRRRCast with HRRR. Both grid-based and object-based metrics indicate HRRRCast performs  better than HRRR at the 20 dBZ threshold.}
\end{figure}

To quantify performance, Figure \ref{obj-metrics} presents object-based verification metrics for composite reflectivity exceeding 20 dBZ. Across metrics such as POD, CSI, FB, and SR, the HRRRCast ensemble consistently outperformed the deterministic HRRR forecast. Notably, HRRRCast achieved higher POD and CSI values, indicating more accurate identification of convective objects, while also maintaining a better success ratio. The bias score for HRRRCast was closer to unity, suggesting improved agreement in the number of detected objects. This preliminary analysis suggests that the object-based verification results are generally consistent with those obtained from grid-based and neighborhood-based approaches.

\begin{figure}
\centering
\includegraphics[width=\linewidth]{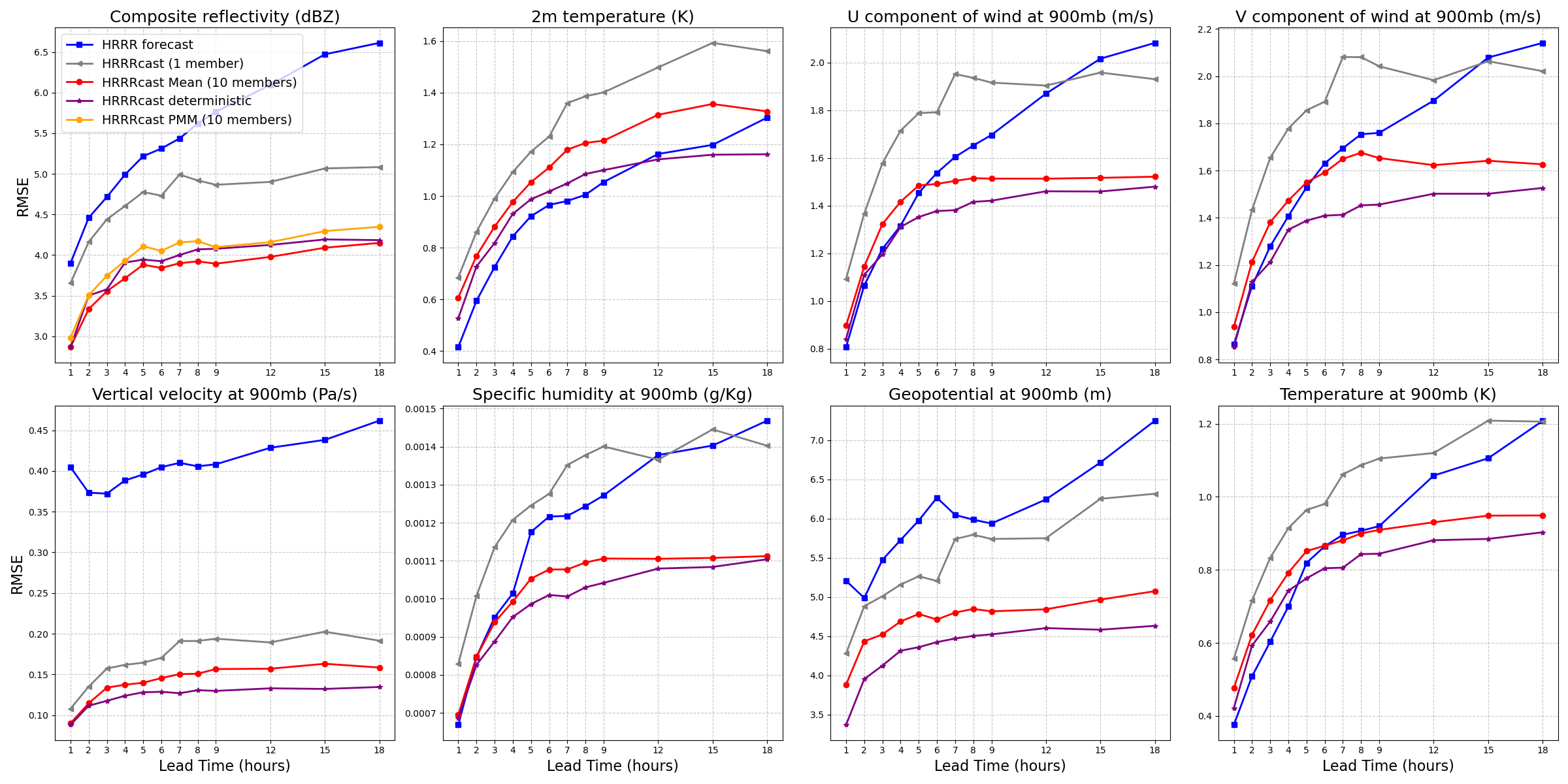}
\caption{Comparison of RMSEs  between the HRRR model and HRRRCast using the HRRR analysis as the ground truth.}
\label{rmse_comp}
\end{figure}

\begin{figure}
\centering
\includegraphics[width=\linewidth]{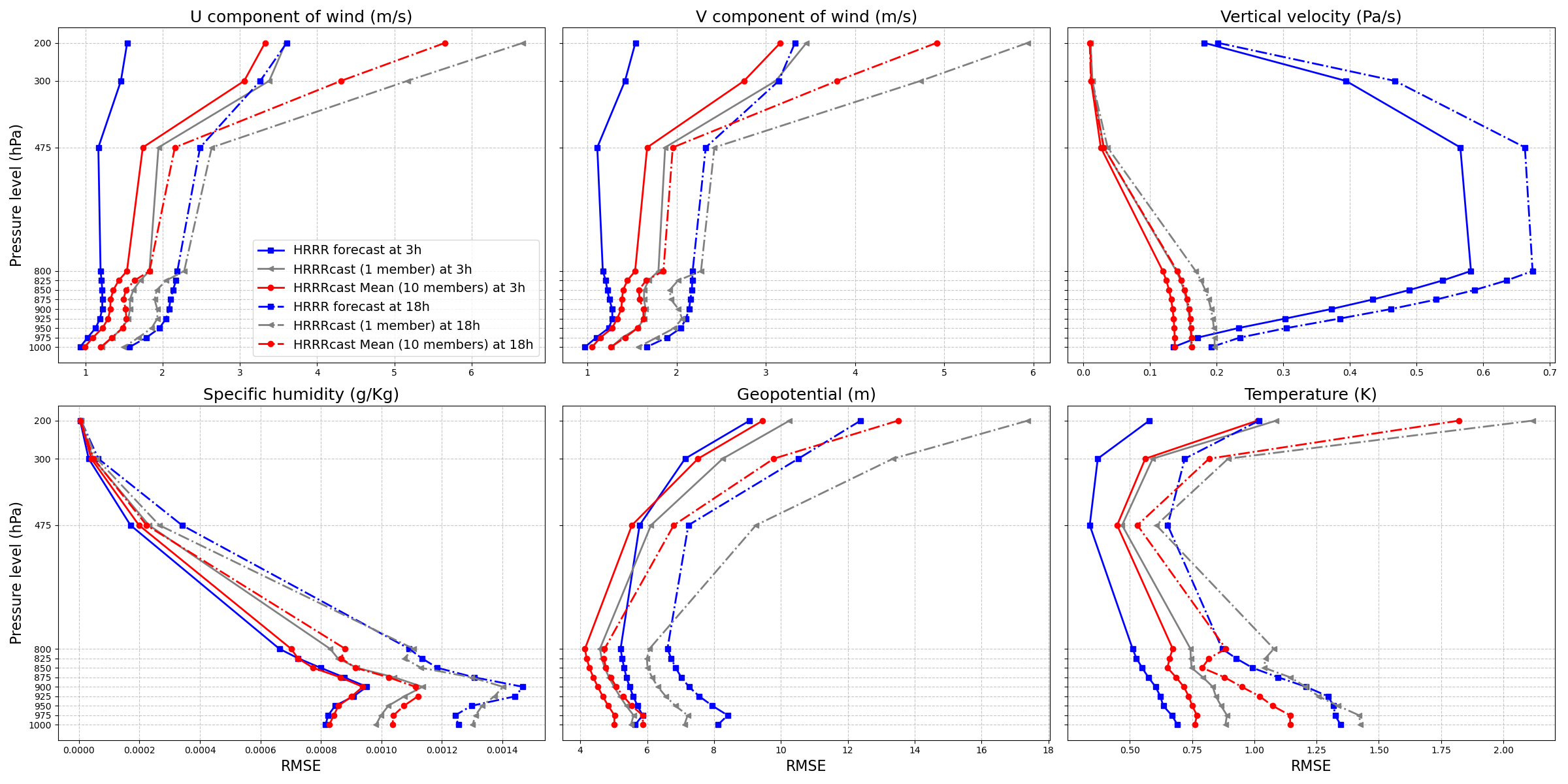}
\caption{Profile of RMSEs for 3D atmospheric variables between HRRR and HRRRCast forecasts of 3h and 18h lead times.}
\label{rmse_comp_profile}
\end{figure}

\subsection{Root mean square errors}
Figure \ref{rmse_comp} compares the RMSEs of selected variables between the single-member HRRRCast, the HRRR model, and the 10-member HRRRCast ensemble, and HRRRCast deterministic. HRRRCast performs particularly well on composite reflectivity, likely due to the higher loss weight (1.0) assigned to the variable during training. In contrast, HRRRCast shows higher RMSEs for 2-meter temperature compared to HRRR, which can be attributed to the lower training weight (0.1) assigned to this variable. For the rest of the variables, except geopotential, the general trend is that HRRR achieves lower RMSEs up to approximately 6 hours, after which HRRRCast ensemble outperforms it.

\begin{figure}
\centering
\includegraphics[width=\linewidth]{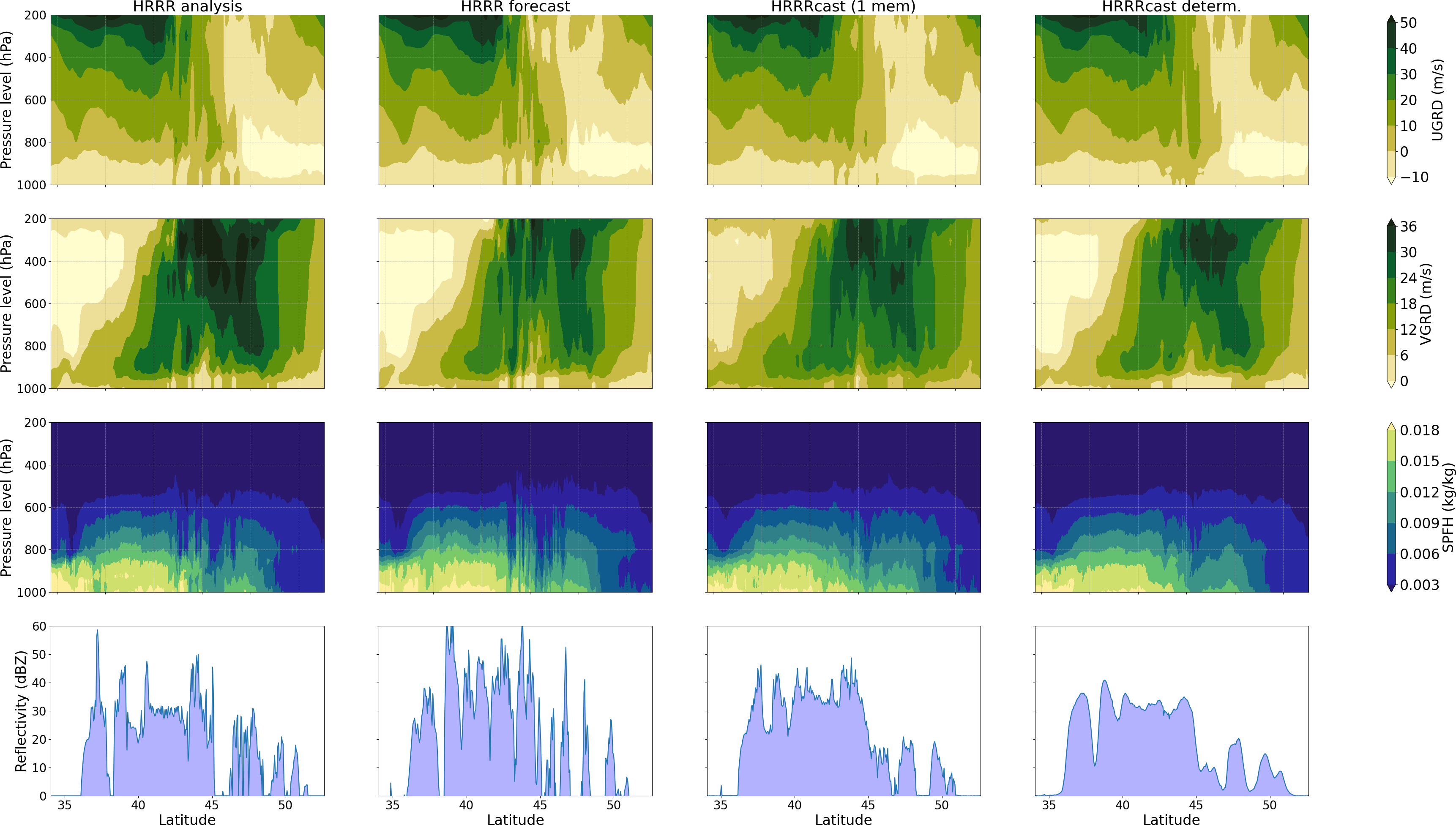}
\caption{Profile of selected atmospheric variables (horizontal wind components and vertical velocity) at central longitude line of the HRRR domain (262.5 W): Forecast initialization time 2024-05-06 23:00 UTC and lead time 1h.}
\label{atmo-profile}
\end{figure}

\begin{figure}
\centering
\includegraphics[width=0.49\linewidth]{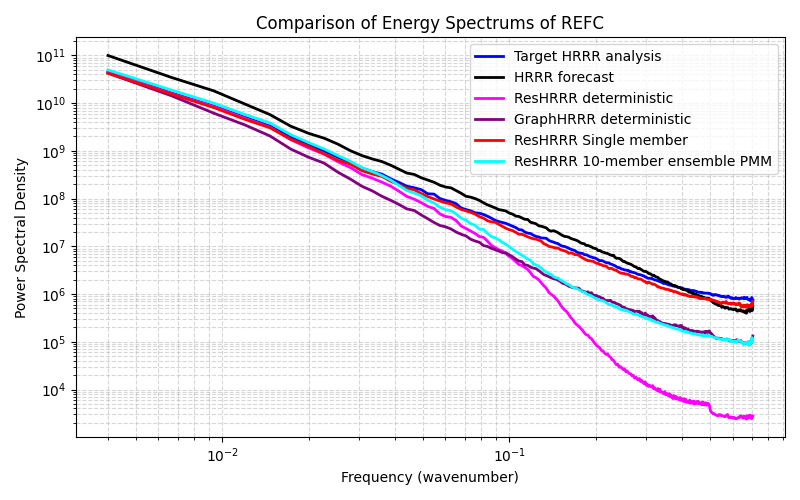}
\includegraphics[width=0.49\linewidth]{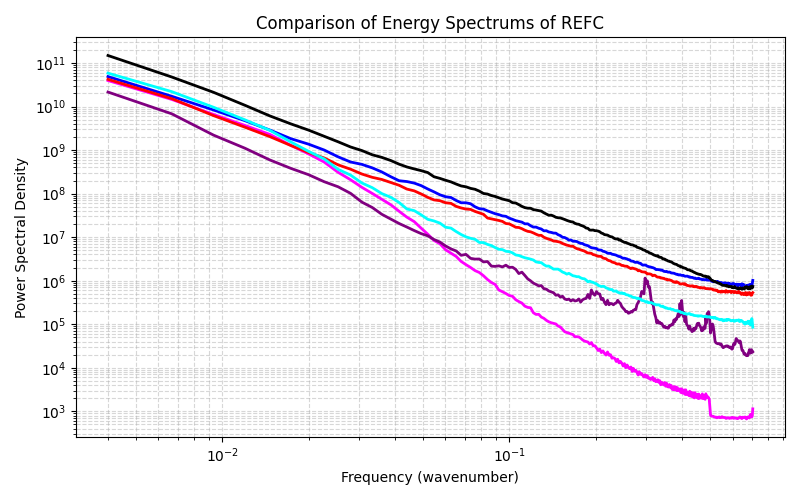}
\includegraphics[width=0.49\linewidth]{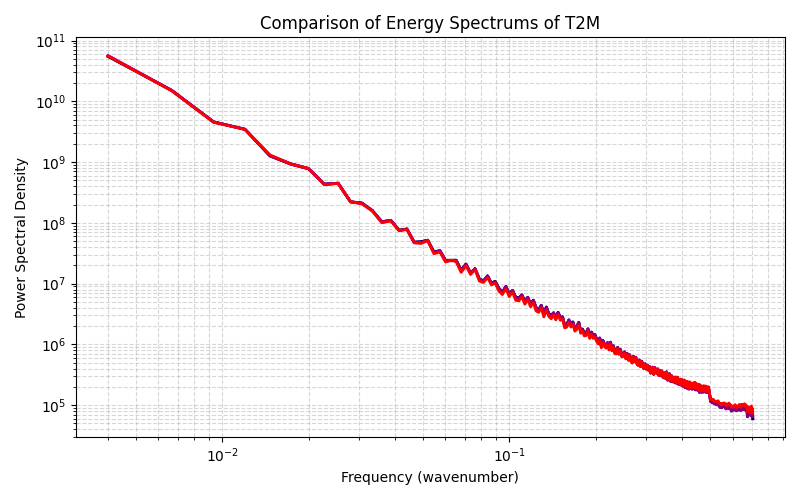}
\includegraphics[width=0.49\linewidth]{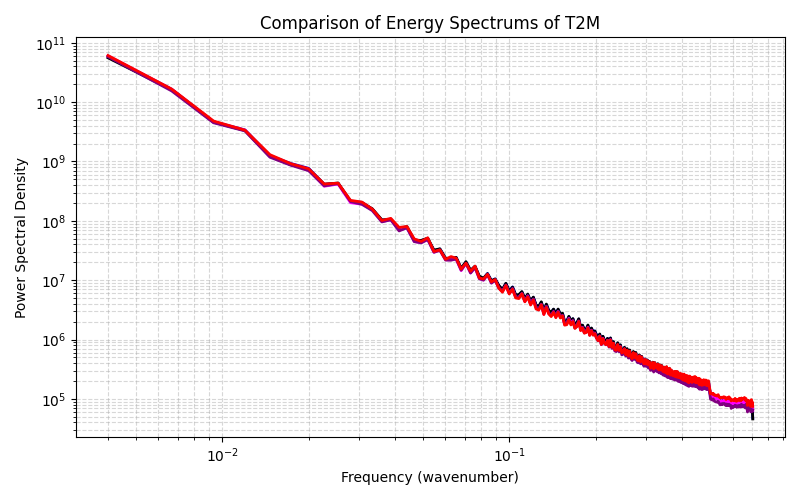}
\caption{Power spectrum comparison of different modesl:  Left) 1h lead time Right) 6h  lead time. Intermittent fields like composite reflectivity (top row) have radically different spectrum than smooth fields like 2m temperature (bottom row).}
\label{power}
\end{figure}

For precipitation-related variables such as composite reflectivity, the appropriate method for computing the ensemble mean is the Probability-Matched Mean (PMM) \citep{clark2017}, thus we also compute the PMM of reflectivity using the method introduced in \citet{potvin2019} (see \ref{append_pmm} for details). As shown in the figure, the PMM ensemble has slightly higher RMSEs than the standard mean, which is expected due to the preservation of intensity distributions. For all other variables, we compute only the standard ensemble mean, which generally results in reduced RMSEs by suppressing high-frequency variability.

Figure \ref{rmse_comp_profile} shows RMSE profiles for several variables across pressure levels at 3-hour and 18-hour lead times. As expected, HRRRCast performs relatively well at lower atmospheric levels (i.e., higher pressures, closer to the surface), aided by the higher loss weight assigned to these layers during training. The ensemble mean outperforms or matches the HRRR forecast across most variables and levels, particularly for geopotential height, and specific humidity and vertical velocity.

Additionally, Figure \ref{atmo-profile} presents vertical profiles of selected atmospheric variables along the mid-longitude transect of the HRRR domain. Despite being limited to 12 pressure levels due to hardware constraints, the vertical structure of several atmospheric variables in the single-member HRRRCast forecast at a 6-hour lead time closely matches the HRRR analysis. The deterministic version of the model also shows good agreement, though wind fields appear somewhat smoothed, reflecting the reduced sharpness often seen in deterministic models.

\subsection{Power spectra}
To assess the spatial sharpness of precipitation forecasts, we compute spatial power spectra across a range of wavenumbers (see Figures \ref{power} and \ref{power-others}), offering a frequency-based evaluation of how well different models capture fine-scale features. The results show substantial differences in the power spectra of composite reflectivity across models, indicating varying abilities to preserve high-frequency spatial structure. In contrast, the power spectra of other variables exhibit relatively minor differences, suggesting that model behavior diverges most significantly in how they represent convective-scale precipitation features.

Deterministic versions of our models significantly damp high-frequency components, even at short lead times. However, GraphHRRR retains slightly more fine-scale content than ResHRRR. Both models are trained with the same mean squared error loss, so this discrepancy may arise from architectural differences — such as graph-based message passing in GraphHRRR versus convolutional processing in ResHRRR — as well as differences in latent representation. GraphHRRR has a larger latent size and a higher overall parameter count (37 million vs. 23 million), which may contribute to its ability to produce sharper results. Differences in input normalization and feature representation may also play a secondary role, potentially influencing how fine-scale features are preserved during training.

The diffusion-based ResHRRR model (HRRRCast) performs notably better in preserving high-frequency spatial features and aligns much more closely with the HRRR analysis data, which was used as its training target. In contrast, the original HRRR model--despite being a high-resolution operational NWP model--exhibits spatial power spectra that deviate significantly from its own analysis fields, indicating some inconsistency between its forecasts and assimilated truths. Notably, HRRRCast maintains sharpness even at 6-hour lead times, while both ResHRRR and GraphHRRR blur substantially. These findings underscore the effectiveness of diffusion models in producing realistic, sharp forecasts and suggest that integrating diffusion modeling with GraphHRRR  could potentially outperform ResHRRR-diffusion by combining architectural strengths with improved generative capabilities.

\subsection{Case study}
We evaluate the performance of HRRRCast during a significant severe weather outbreak from May 6--10, 2024 \citep{nws2024may7}, focusing on a squall line event centered on May 7.

\begin{figure}
\centering
\fbox{\includegraphics[width=0.47\linewidth]{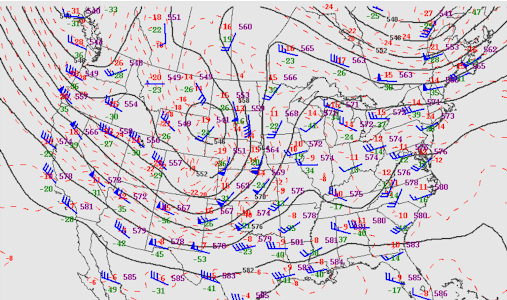}}
\fbox{\includegraphics[width=0.49\linewidth]{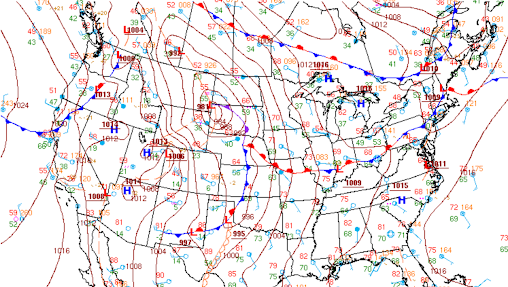}}
\caption{500mb geopotential and surface analysis map for our case study on May 07, 2024 00:00 UTC.}
\label{case_study_maps}
\end{figure}

Figure \ref{case_study_maps} shows maps from the Weather Prediction Center (WPC) archive. At 0000 UTC, the 500 mb geopotential height field showed a well-defined upper-level low centered over the Dakotas. This feature contributed to strong northwesterly winds aloft across the western regions and south-southwesterly flow over the central Plains. At the surface, a deepening low-pressure system tracked eastward from Wyoming into the South Dakota region, reaching a minimum central pressure of 980 mb by the early morning of May 8, 2024. The presence of warm, moist air near the surface, combined with sufficient wind shear aloft, created favorable conditions for the development of severe thunderstorms.

At 0000 UTC, a partially organized convective line was already active, extending north to south across the eastern portions of South Dakota, Nebraska, and northern Kansas. A strong, multi-cell storm oriented southwest to northeast was also present over Oklahoma as an extension of this convective line. As time progressed, the system moved eastward and became more organized, developing into a well-defined north-south oriented squall line with a clear leading edge by 0600 UTC.

\begin{figure}
\centering

\noindent
\makebox[0.22\linewidth][c]{HRRR analysis}%
\hfill
\makebox[0.22\linewidth][c]{HRRR forecast}%
\hfill
\makebox[0.22\linewidth][c]{HRRRCast 1-mem.}%
\hfill
\makebox[0.28\linewidth][l]{HRRRCast  determ.}%

\vspace{0.5em}
\includegraphics[width=0.97\linewidth]{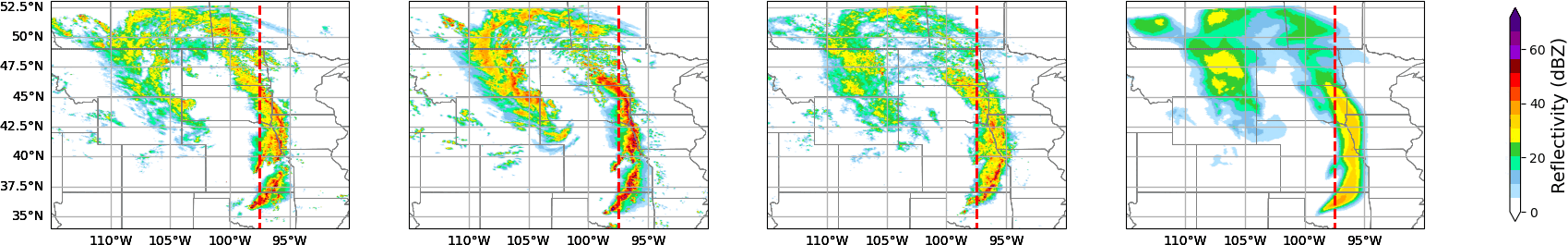}\\
\vspace{0.5em}
\includegraphics[width=0.97\linewidth]{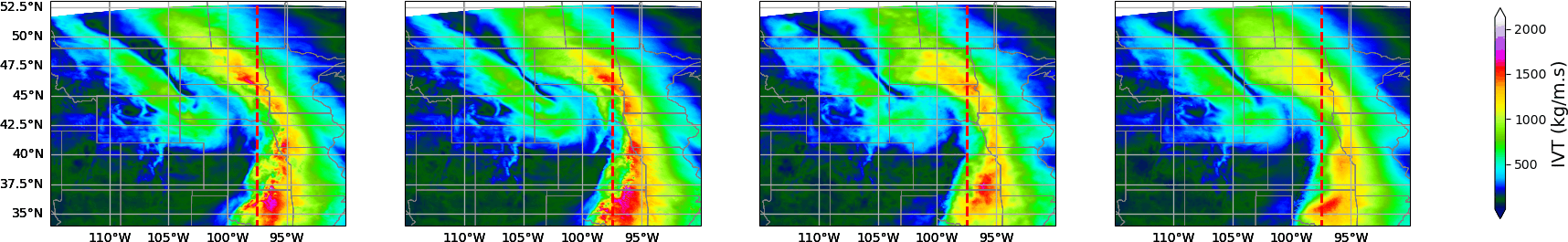} \\
\vspace{0.5em}
\includegraphics[width=0.97\linewidth]{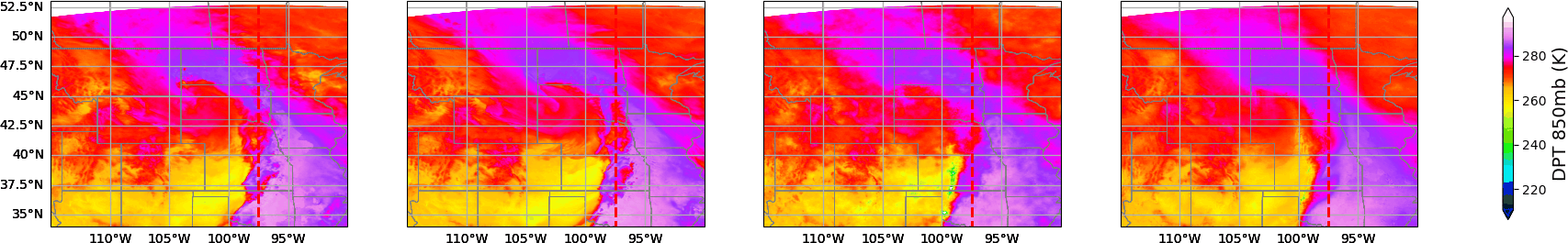} \\
\vspace{0.5em}
\includegraphics[width=1.0\linewidth]{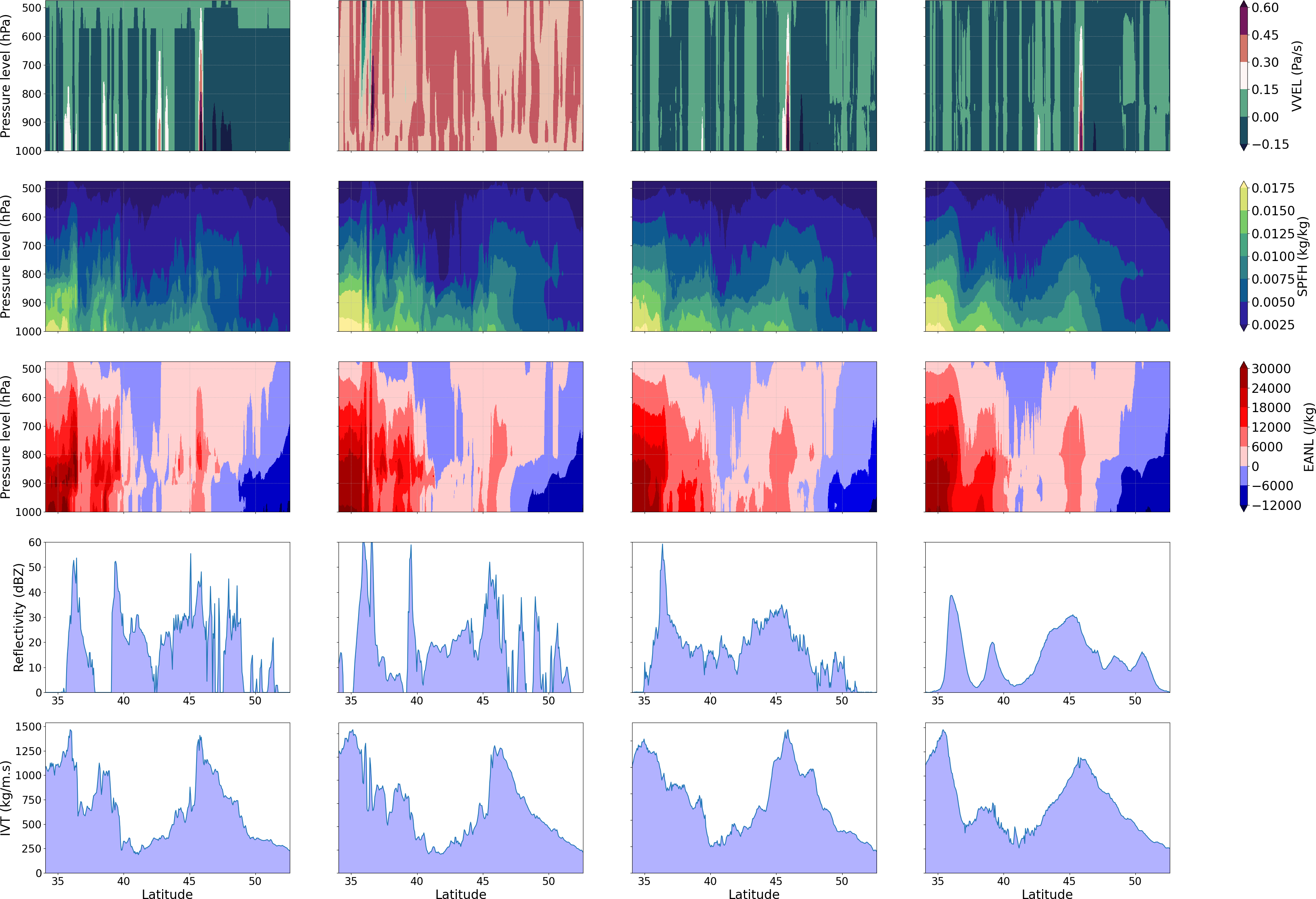} \\

\caption{Case study: A 3h forecast made by different models at forecast initialization time: 2024-05-06 23:00 UTC. Quantities such as Integrated Vapor Transport (IVT),
Dew Point Temperature (DPT) are derived from the variables modeled in HRRRCast. Additionally, the profile of specific humidity, vertical velocity and Enthalpy Anomaly across a vertical 
section (red line) are shown. High intensity radar reflectivity  values coincide with high values of corresponding IVT and DPT values for almost all models. }
\label{case_study_3h}

\end{figure}

Figure \ref{case_study_3h} presents composite reflectivity, Integrated Vapor Transport (IVT), 850 mb dew point temperature, and vertical velocity over a smaller region in the Upper Midwest. These fields are shown for the HRRR analysis, HRRR forecast, a single HRRRCast ensemble member, and the deterministic HRRRCast forecast (from left to right), all for a 3-hour lead time based on model runs initialized at 2300 UTC on May 6. Additionally, vertical cross sections along $100^\circ \text{W}$ longitude are provided, displaying vertical velocity, specific humidity, enthalpy anomaly, IVT, and composite reflectivity.

Focusing on the two-dimensional composite reflectivity in the top row, the HRRR forecast generally aligns well with the HRRR analysis in terms of overall structure and spatial coverage. However, it notably overestimates reflectivity intensity along the leading edge of the squall line. It also overpredicts both the intensity and areal extent of reflectivity in the ``wraparound" region over Montana and Wyoming. In contrast, the selected HRRRCast ensemble member slightly underestimates reflectivity overall compared to the HRRR analysis but offers a somewhat better representation of system features, especially in the wraparound area. This trend is confirmed in Figures \ref{case_study_1h} and \ref{case_study_6h}. 

As expected, the deterministic HRRRCast output appears smoother, particularly at 6h lead time (Fig. \ref{case_study_6h}), and captures only the general location of the convective system -- highlighting the limitations of deterministic AI models for convection-allowing modelling.

Differences in composite reflectivity between the HRRR and HRRRCast forecasts are accompanied by an overall overprediction of IVT and an underprediction of 850 mb dew point temperature, relative to the HRRR analysis. Vertical velocity cross sections reveal significant differences between the HRRR forecast and other solutions. In the HRRR forecast, there is an almost continuous updraft region, varying in strength. Conversely, both the HRRRCast ensemble and deterministic forecasts display isolated strong updrafts interspersed with zones of weak vertical motion or downdrafts. This pattern more closely matches the HRRR analysis, although both HRRRCast forecasts slightly underestimate the strength of some weaker updrafts at lower latitudes.
For specific humidity, both HRRR and HRRRCast forecasts depict elevated moisture levels higher in the atmosphere compared to the HRRR analysis, particularly at lower latitudes. Enthalpy anomaly fields are similar across all forecasts, showing increased heat content extending further aloft relative to the analysis, again most notably in lower latitudes. The vertical IVT cross section exhibits consistent behavior across all solutions, while the composite reflectivity cross section shows a slightly better match between the HRRR forecast and HRRR analysis than with the HRRRCast outputs.

Overall, the case study underscores HRRRCast's strengths in avoiding the overprediction bias of HRRR, capturing storm intensity with greater realism, and providing credible vertical motion patterns. These features are critical for severe weather forecasting and point to the value of HRRRCast as a data-driven complement to HRRR, especially for ensemble-based applications.

\section{Conclusions}
We have developed HRRRCast, a data-driven emulator for the High-Resolution Rapid Refresh (HRRR) model, designed to support regional weather forecasting at convection-allowing scales. HRRRCast builds upon and extends the capabilities of the StormCast framework by introducing key advancements: full CONUS coverage, multi-lead time training (1h, 3h, 6h) in a single model, the use of HRRR analysis data instead of forecasts for training, and the incorporation of future GFS states to guide both forecasting and downscaling. While both HRRR and StormCast operate at 3-km grid spacing, HRRRCast uses a coarser 6-km grid to enable full CONUS coverage within hardware constraints -- a tradeoff we believe is justified by the ability to train on storm activity across the entire U.S.

A major strength of HRRRCast is its ability to generate probabilistic ensemble forecasts via diffusion modeling. Using the DDIM, HRRRCast can produce ensembles that quantify forecast uncertainty while remaining far more computationally efficient than traditional physics-based ensembles. This opens the door to cost-effective ensemble forecasting for HRRR, addressing limitations of expensive deterministic modeling and time-lagged ensemble approaches currently used operationally (e.g., HRRR-TLE). HRRRCast thus has the potential to augment or even replace existing HRRR systems, especially in ensemble applications.

We evaluate HRRRCast using three complementary verification approaches: grid-based, neighborhood-based (FSS), and object-based methods. Our findings demonstrate that HRRRCast outperforms HRRR at the 20 dBZ threshold across all lead times, and achieves comparable performance at 30 dBZ. Furthermore, the model maintains skill at longer lead times due to the persistent benefits of GFS-guided downscaling, which helps stabilize forecasts beyond 9 hours.

The ResHRRR architecture powering HRRRCast integrates several novel techniques, including squeeze-and-excitation (SE) blocks, FiLM conditioning on lead time and diffusion step, and trainable skip connections. Together with DDIM-based sampling, these design choices help preserve high-frequency spatial detail and reduce the blurriness typically seen in deterministic models. HRRRCast's probabilistic forecasts also better balance detection and false alarm rates compared to HRRR, showing reduced bias and more calibrated uncertainty.

While promising, limitations remain. The current GraphHRRR variant underperforms due to the absence of diffusion modeling, lack of synoptic-scale inputs, and insufficient training for long lead times. Future development will focus on integrating GenCast-style diffusion models with graph neural networks, combining geometric flexibility with improved generative fidelity. Additionally, ensemble spread calibration remains an open challenge, which we aim to improve via more diverse noise perturbations and initial condition perturbations using external systems like GEFS.

In summary, HRRRCast demonstrates the viability of data-driven emulation of operational regional NWP systems, offering both deterministic and ensemble capabilities. Its strengths in cost-effective ensemble generation, forecast sharpness, and downscaling skill make it a valuable complement to physics-based systems and a potential path forward for operational AI-enhanced regional weather prediction.

\section*{Acknowledgment}
 The authors gratefully acknowledge funding from the Software Environment for  Novel Architectures (SENA) effort at GSL. This research used resources of the National Energy Research Scientific Computing Center, a DOE Office of Science User Facility supported by the Office of Science of the U.S. Department of Energy under Contract No. DE-AC02-05CH11231 using NERSC award GenAI@NERSC DDR-ERCAP0034078. We would like to thank Ligia Bernardet for her support and providing  feedback on the work, Jeff Beck and Gerard Ketefian for help with verification with MET, and members of the ``NOAA AI4NWP forum" for interesting discussions.

\bibliographystyle{elsarticle-harv}
\bibliography{references,mypublications}

\clearpage
\appendix

\section{Ensemble calibration}
\label{append_spread}
\begin{figure}[H]
\centering
\includegraphics[width=0.7\linewidth]{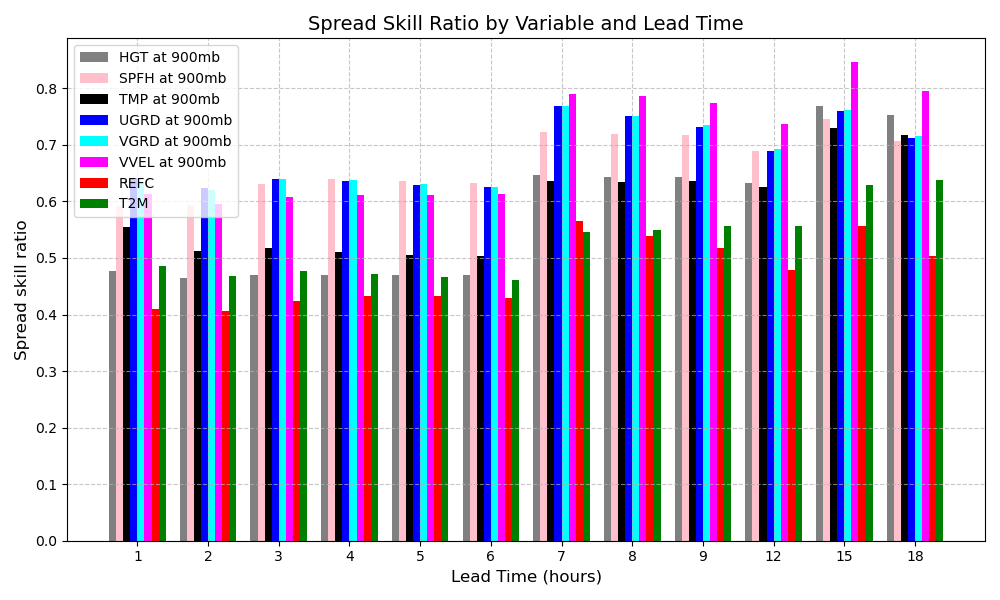}
\caption{Spread-error ratio.}
\label{ser}
\end{figure}

The spread-error ratio provides insight into a model's ensemble reliability and calibration. The spread-error ratio compares the ensemble spread (standard deviation across members) to the actual forecast error (RMSE between ensemble mean and observations/ground truth). A well-calibrated ensemble should have a spread-error ratio near 1.0, indicating that the ensemble spread accurately represents forecast uncertainty.
Figure \ref{ser} shows the spread-error ratio of HRRRCast on selected variables is on average 0.6, suggesting that the ensemble is under-dispersive -- the spread is smaller than the actual forecast errors. Looking ahead, we plan to improve this by incorporating GEFS-based initial condition perturbations rather than relying solely on noise perturbations in the diffusion process, and by adding stochastic noise during DDIM sampling steps rather than using deterministic sampling. These approaches would better represent both initial condition uncertainty and model uncertainty, leading to more realistic ensemble spread.

\section{Probability Matched Mean}
\label{append_pmm}
To construct an ensemble mean for composite reflectivity that preserves both the spatial coherence and realistic intensity distribution of individual ensemble members, we use the Probability Matched Mean (PMM) approach following \citet{potvin2019}. The method proceeds in three steps:
\begin{enumerate}
\item Compute the ensemble mean field by averaging the reflectivity across all ensemble members at each grid point. This provides the spatial structure that will be preserved in the final PMM field.
\item Pool and sort all reflectivity values from all ensemble members into a single 1D array. From this sorted array, every $N^{th}$ value is selected -- where $N$ is the number of ensemble members -- to produce a downsampled set of values. This set approximates the typical value distribution across the ensemble without biasing toward any single member.
\item Reassign the sorted values from step 2 to the grid points of the mean field based on rank ordering: the largest value is assigned to the location with the highest mean, the second-largest to the second-highest, and so on. This ensures the final field has the same spatial pattern as the ensemble mean, but with a value distribution that better reflects the ensemble variability.
\end{enumerate}
This method captures realistic high-intensity features that a traditional ensemble mean would suppress due to averaging, while avoiding the need to sort each member individually.

\section{Extended results}
\label{append_extended}

\begin{figure}[H]
\centering
\includegraphics[width=0.49\linewidth]{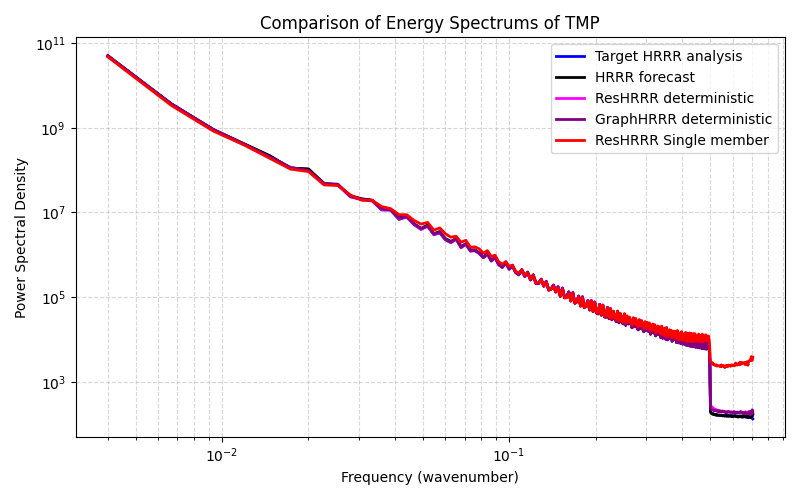}
\includegraphics[width=0.49\linewidth]{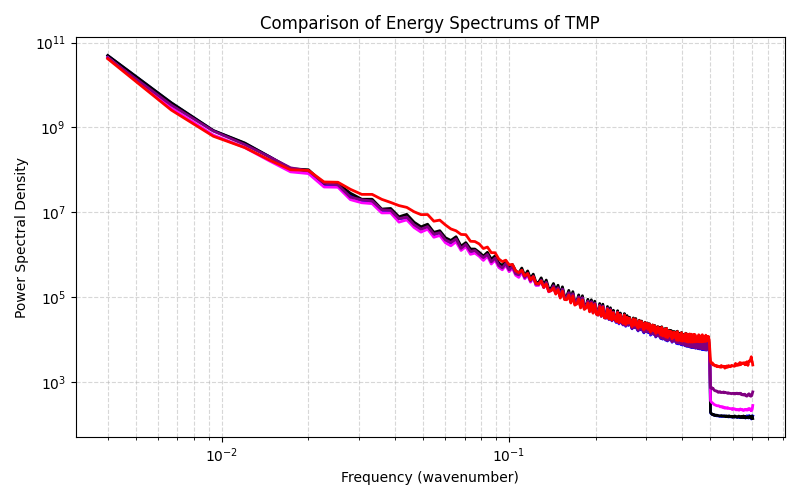}
\includegraphics[width=0.49\linewidth]{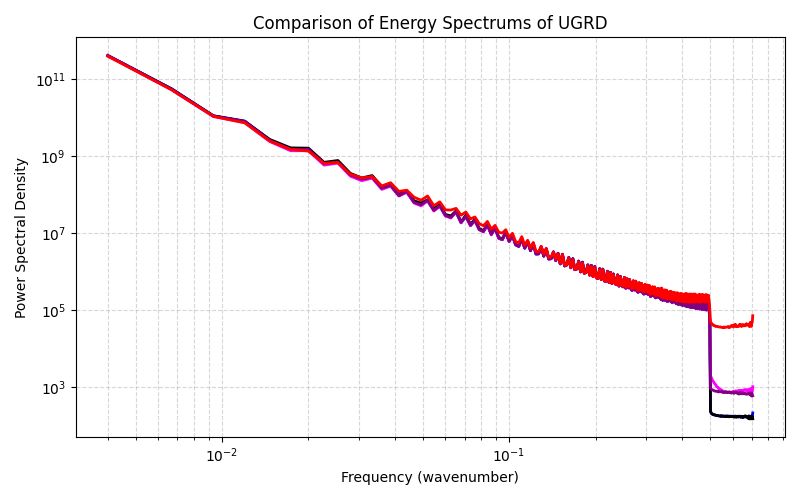}
\includegraphics[width=0.49\linewidth]{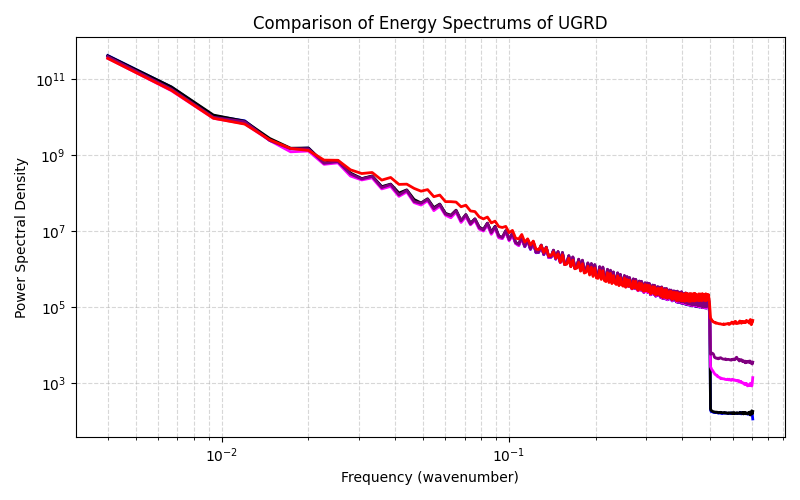}
\includegraphics[width=0.49\linewidth]{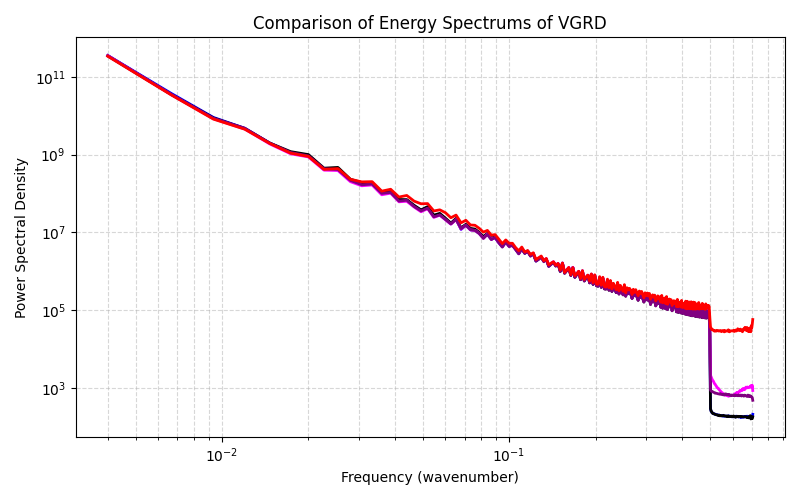}
\includegraphics[width=0.49\linewidth]{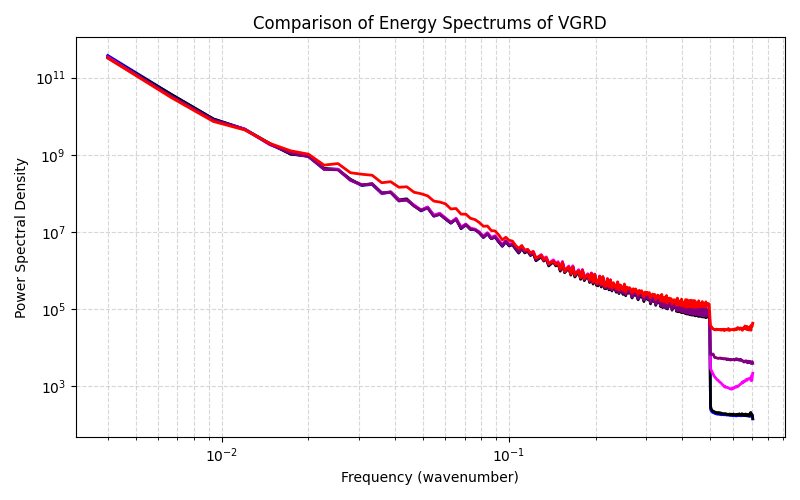}
\includegraphics[width=0.49\linewidth]{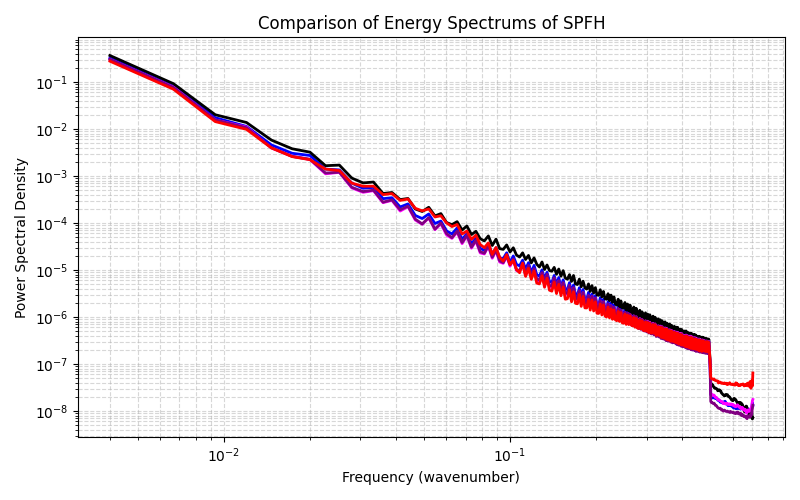}
\includegraphics[width=0.49\linewidth]{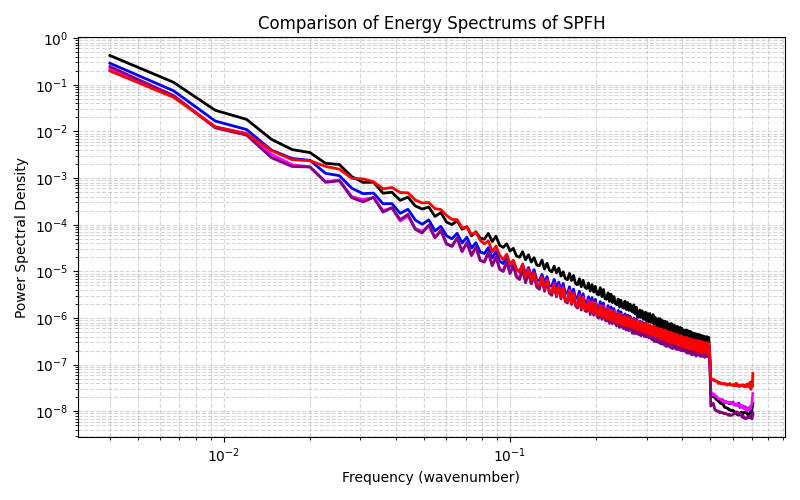}
\caption{Power spectrum comparison of HRRRCast deterministic and probabilistic models with HRRR analysis and HRRR.  Left) 1h lead time Right) 6h  lead time.  Power spectrum of
 these relatively smooth fields  do not show significant differences between different modes.}
\label{power-others}
\end{figure}

\begin{figure}
\centering
\includegraphics[width=0.327\linewidth]{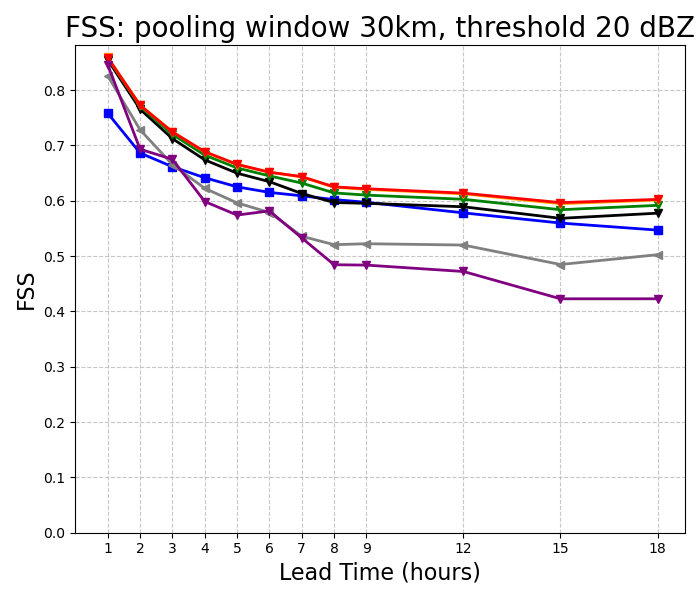}
\includegraphics[width=0.327\linewidth]{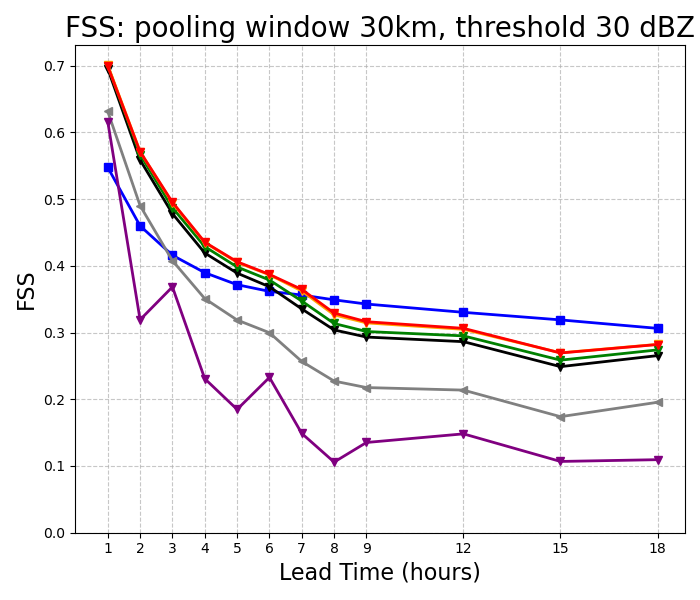}
\includegraphics[width=0.327\linewidth]{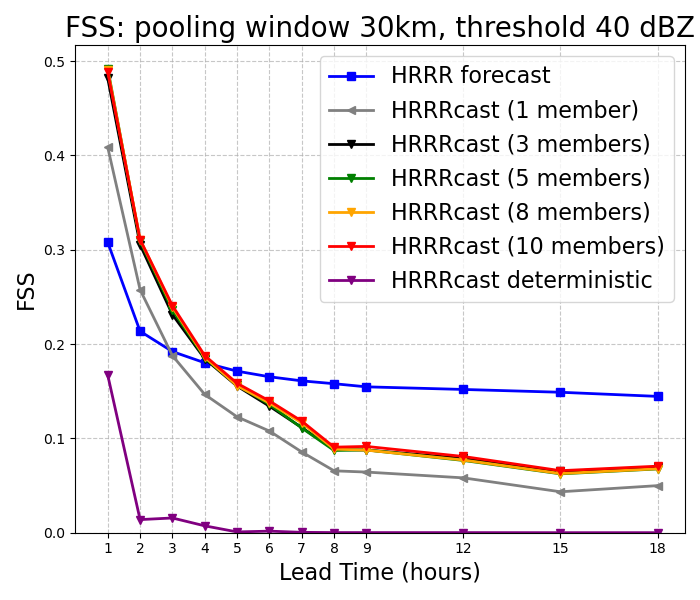}
\includegraphics[width=0.327\linewidth]{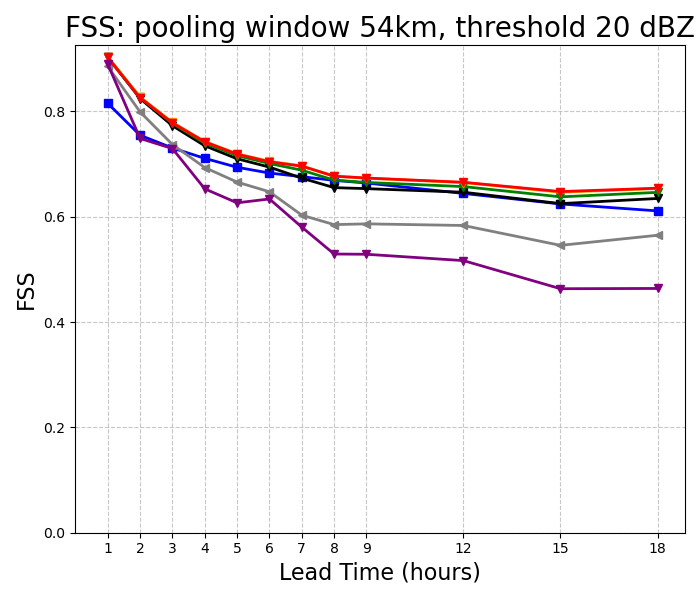}
\includegraphics[width=0.327\linewidth]{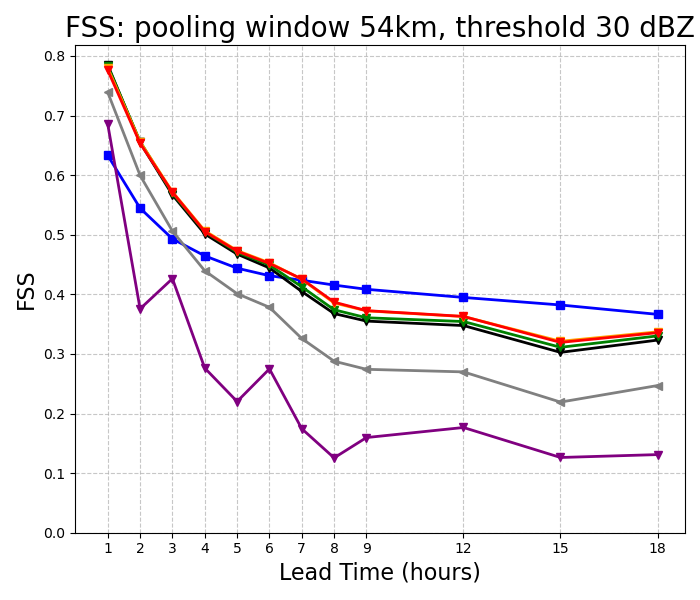}
\includegraphics[width=0.327\linewidth]{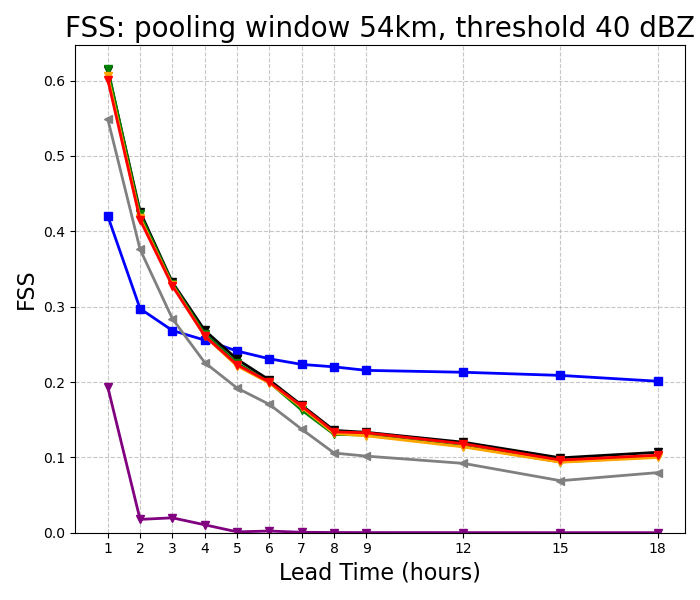}
\includegraphics[width=0.327\linewidth]{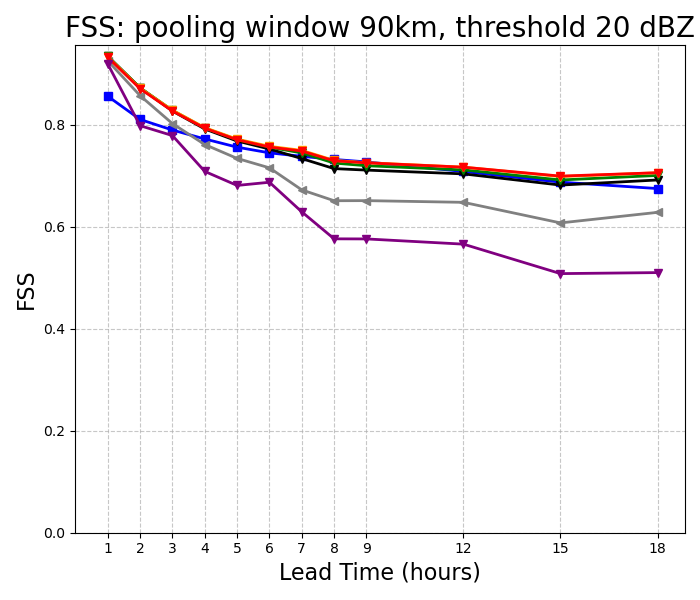}
\includegraphics[width=0.327\linewidth]{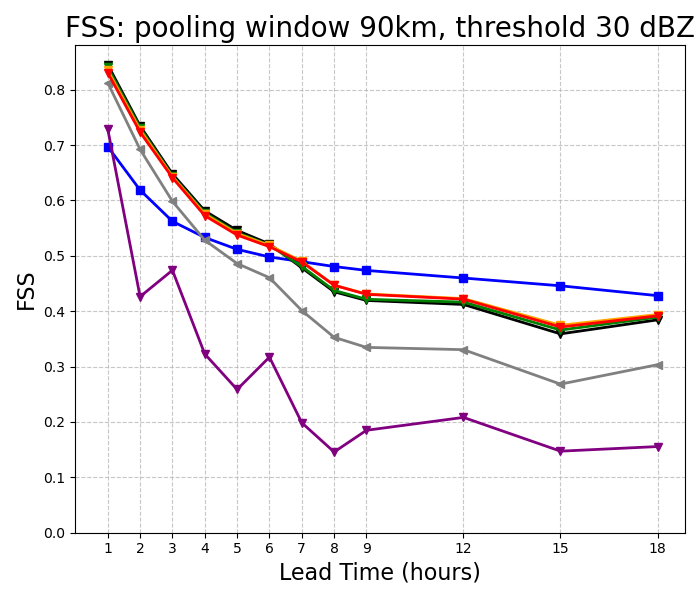}
\includegraphics[width=0.327\linewidth]{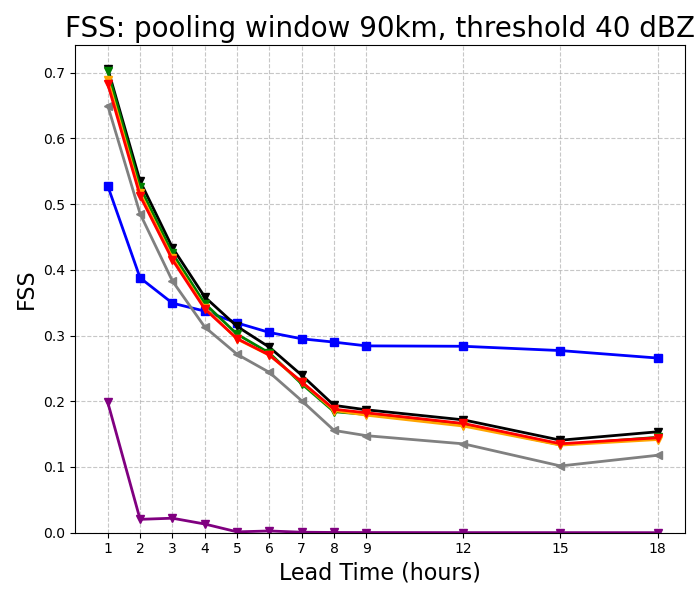}
\caption{Fractions Skill Score (FSS) of composite reflectivity computed over the 10 days dataset at hourly interval. Results for different dBZ thresholds (20 dBZ, 30 dBZ, and 40 dBZ) and pooling windows (30km, 54km and 90km) are shown.}
\label{fss-others}
\end{figure}

\begin{figure}
\centering
\includegraphics[width=0.327\linewidth]{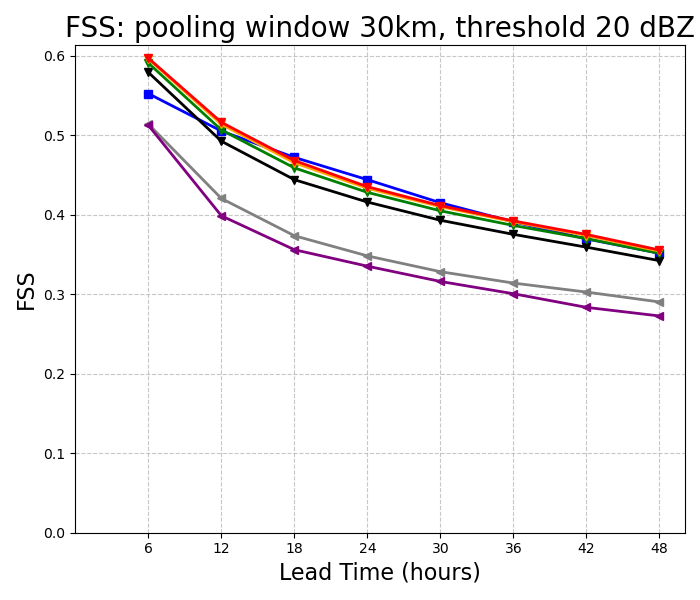}
\includegraphics[width=0.327\linewidth]{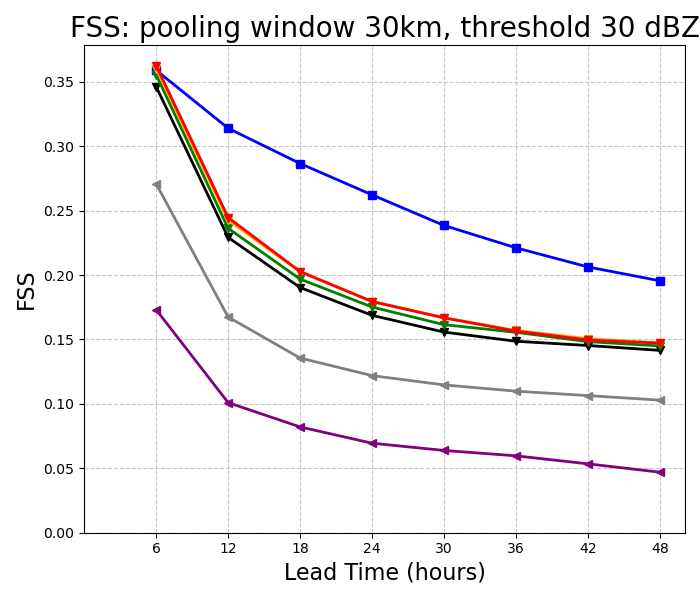}
\includegraphics[width=0.327\linewidth]{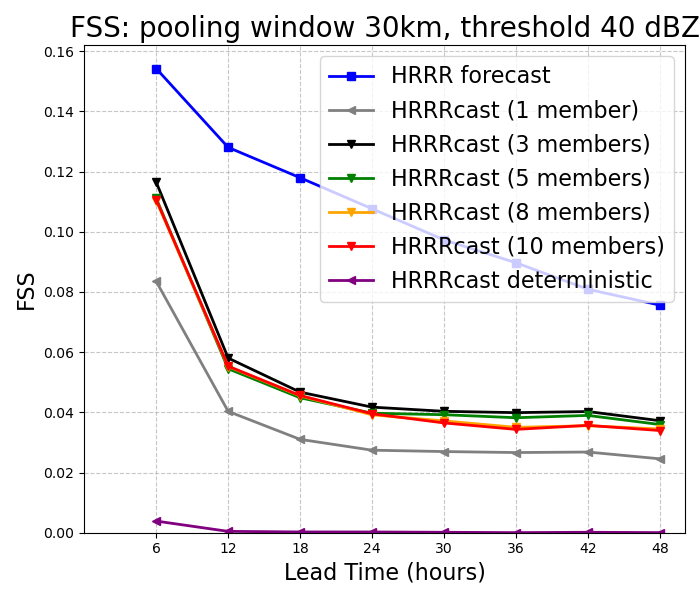}
\includegraphics[width=0.327\linewidth]{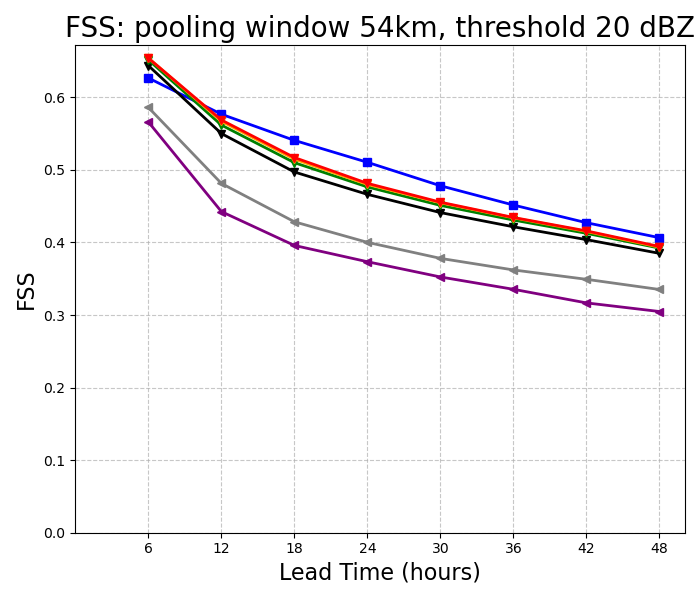}
\includegraphics[width=0.327\linewidth]{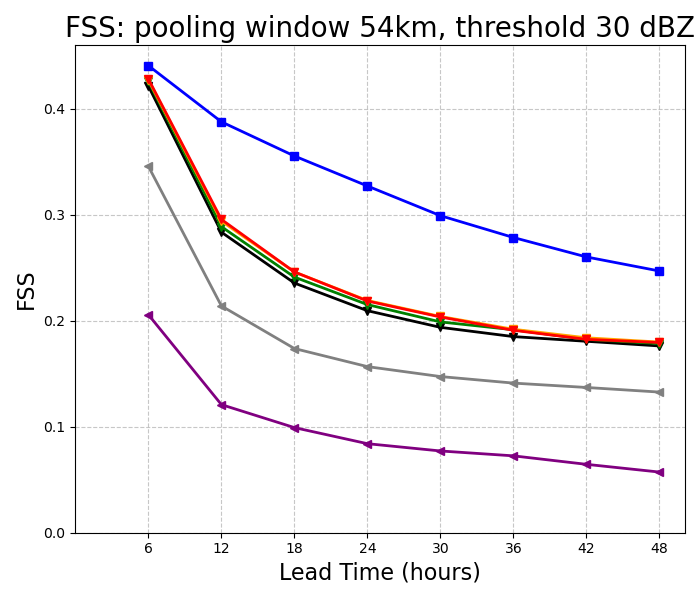}
\includegraphics[width=0.327\linewidth]{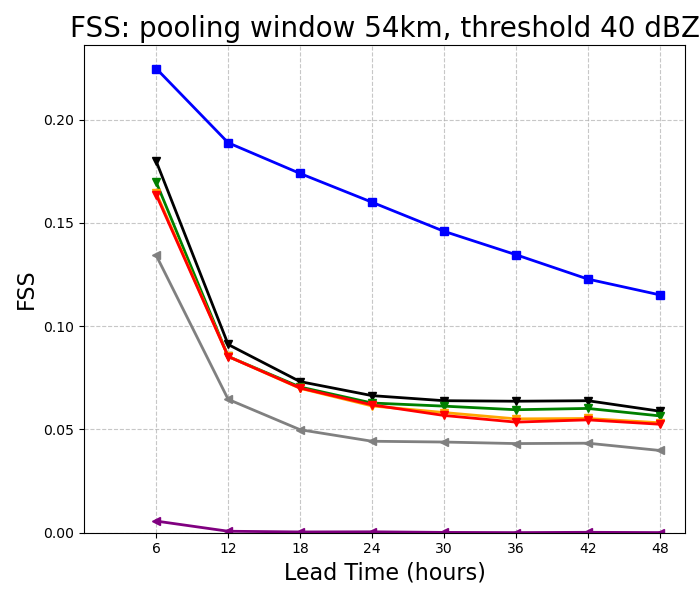}
\includegraphics[width=0.327\linewidth]{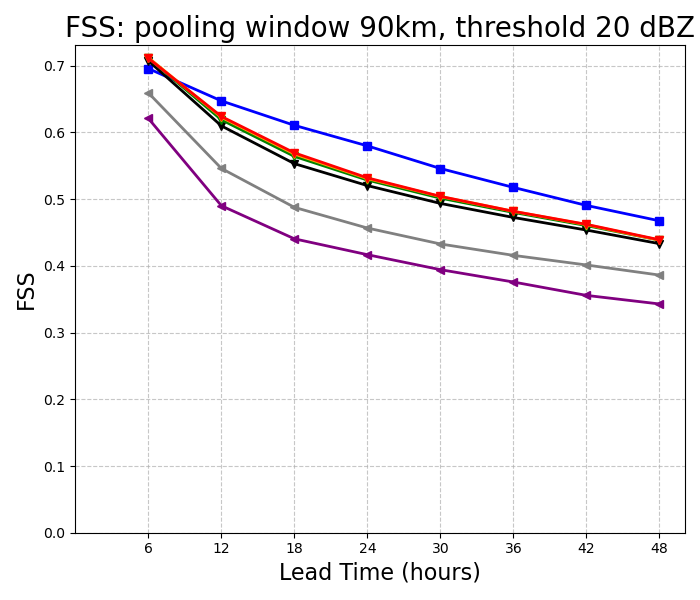}
\includegraphics[width=0.327\linewidth]{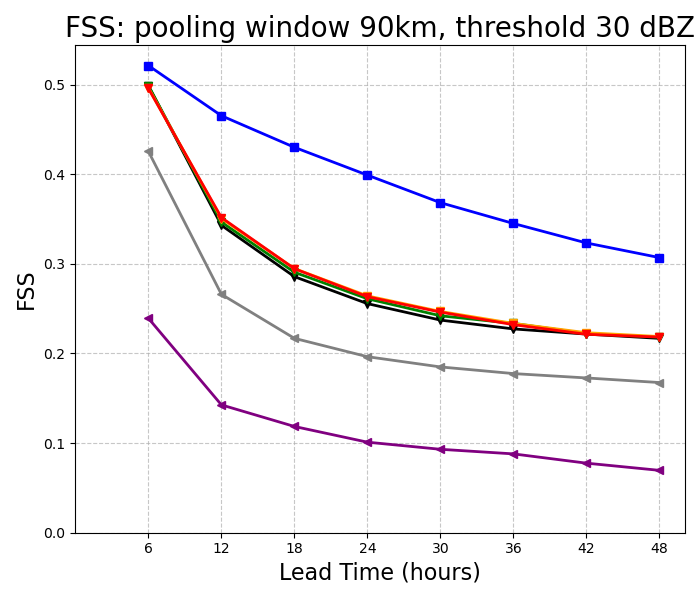}
\includegraphics[width=0.327\linewidth]{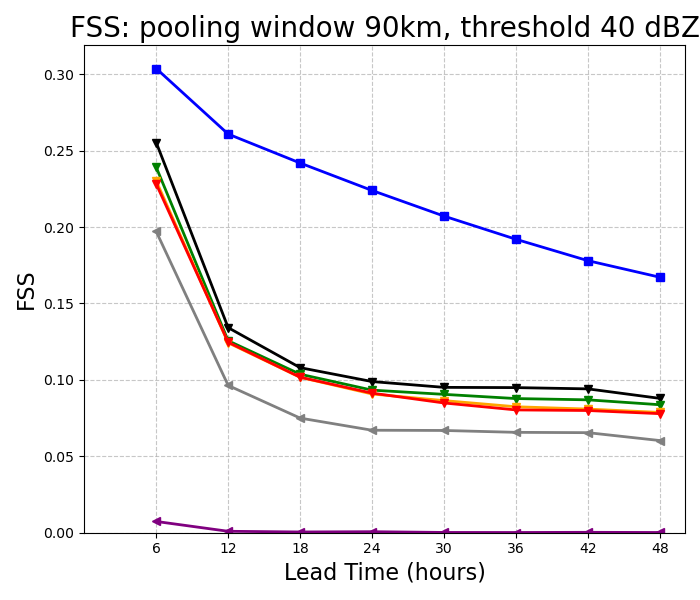}
\caption{Fractions Skill Score (FSS) of composite reflectivity computed over the 4-month dataset at 6-hourly interval. Results for different dBZ thresholds (20 dBZ, 30 dBZ, and 40 dBZ) and pooling windows (30km, 54km and 90km) are shown.}
\label{fss-others-4month}
\end{figure}

\begin{figure}
\centering

\noindent
\makebox[0.22\linewidth][c]{HRRR analysis}%
\hfill
\makebox[0.22\linewidth][c]{HRRR forecast}%
\hfill
\makebox[0.22\linewidth][c]{HRRRCast 5-mem.}%
\hfill
\makebox[0.24\linewidth][c]{HRRRCast 1-mem.}%

\vspace{0.5em}

\setlength{\tabcolsep}{4pt} 
\begin{tabular}{@{} l c @{}}
  \includegraphics[width=\linewidth]{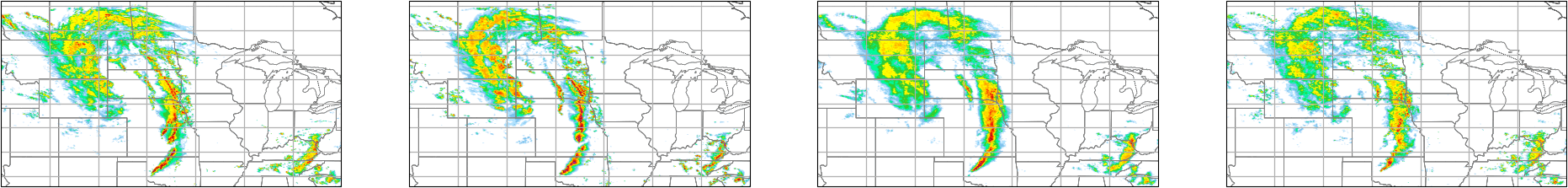} & f01 \\[6pt]
  \includegraphics[width=\linewidth]{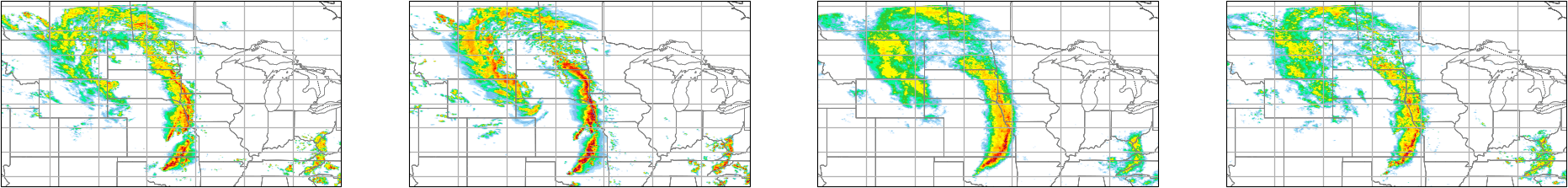} & f03 \\[6pt]
  \includegraphics[width=\linewidth]{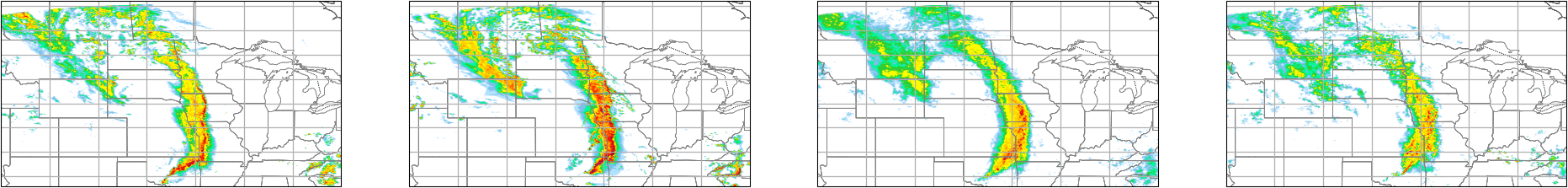} & f06 \\[6pt]
  \includegraphics[width=\linewidth]{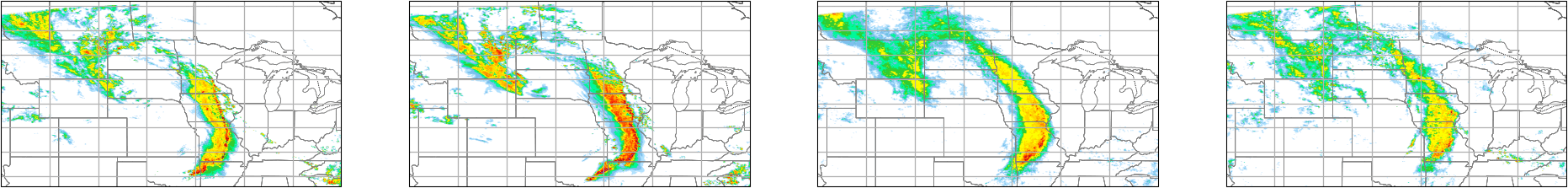} & f09 \\[6pt]
  \includegraphics[width=\linewidth]{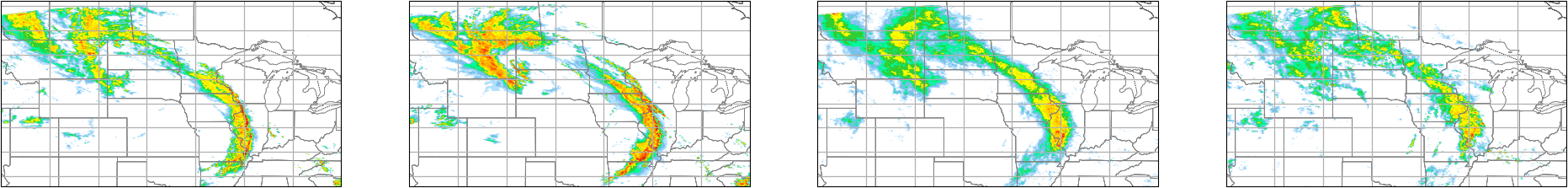} & f12 \\[6pt]
  \includegraphics[width=\linewidth]{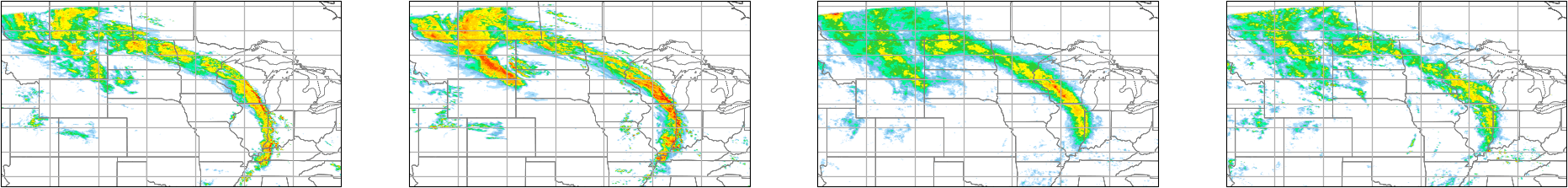} & f15 \\[6pt]
  \includegraphics[width=\linewidth]{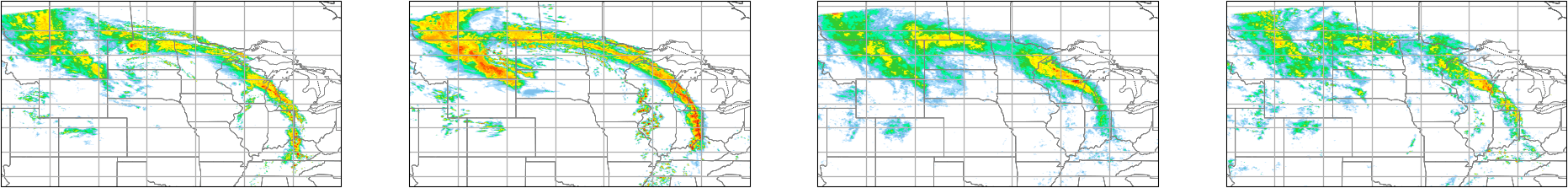} & f18 \\ [6pt]
\multicolumn{2}{c}{
  \begin{minipage}{0.4\linewidth}
    \centering
    \includegraphics[width=\linewidth]{pics/refc-colorbar.png} \\
    \small Composite reflectivity (dBZ)
  \end{minipage}
} \\
\end{tabular}

\caption{Enlarged view of composite reflectivity: forecast initialization time 2024-05-06 23:00 UTC}
\label{refc_map2}

\end{figure}

\begin{figure}
\centering

\noindent
\makebox[0.22\linewidth][c]{HRRR analysis}%
\hfill
\makebox[0.22\linewidth][c]{HRRR forecast}%
\hfill
\makebox[0.22\linewidth][c]{HRRRCast 5-mem.}%
\hfill
\makebox[0.24\linewidth][c]{HRRRCast 1-mem.}%

\vspace{0.5em}

\setlength{\tabcolsep}{4pt} 
\begin{tabular}{@{} l c @{}}
  \includegraphics[width=\linewidth]{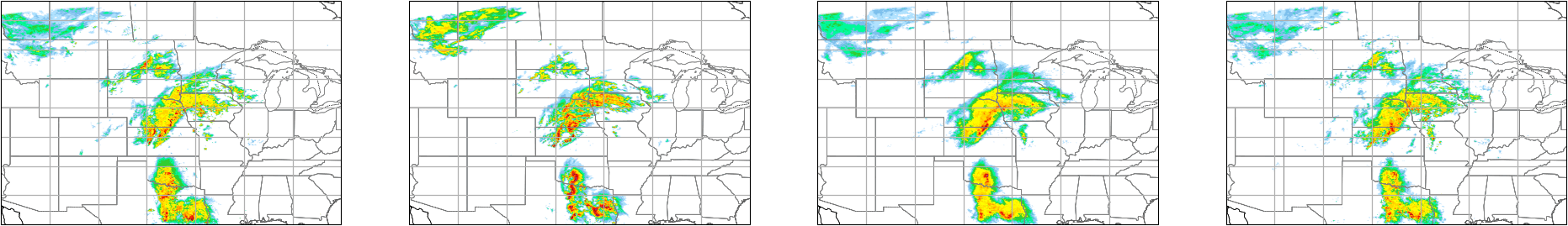} & f01 \\[6pt]
  \includegraphics[width=\linewidth]{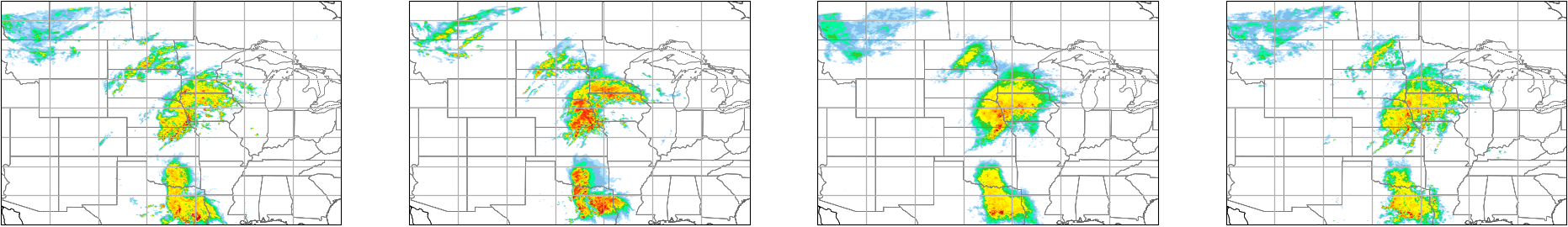} & f03 \\[6pt]
  \includegraphics[width=\linewidth]{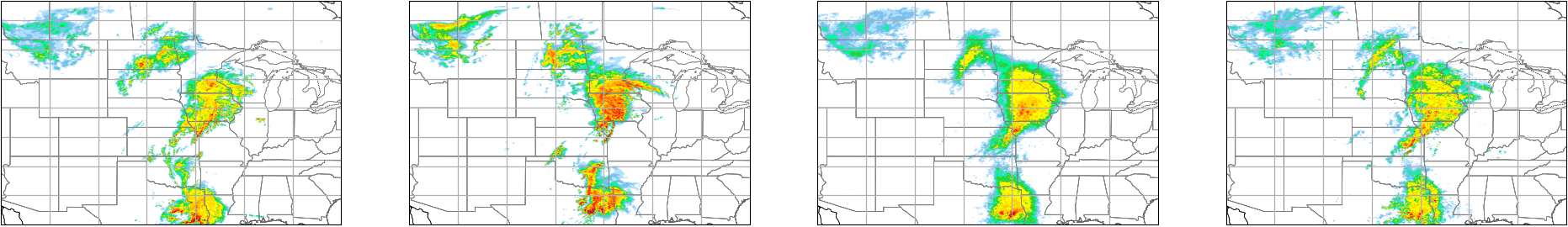} & f06 \\[6pt]
  \includegraphics[width=\linewidth]{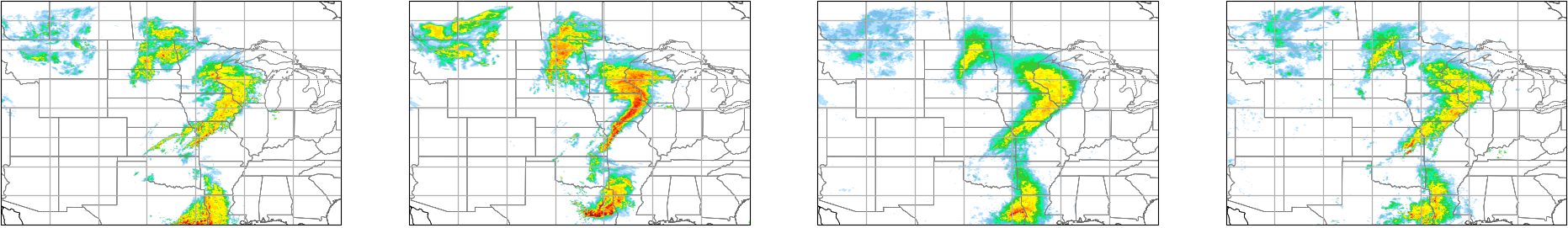} & f09 \\[6pt]
  \includegraphics[width=\linewidth]{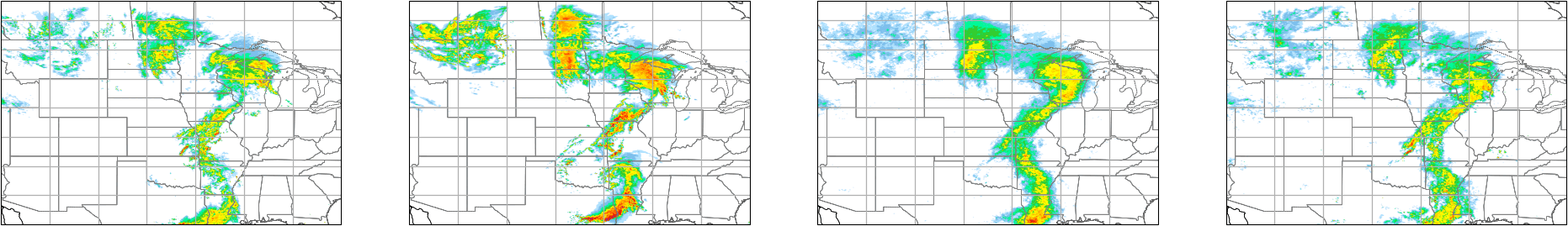} & f12 \\[6pt]
  \includegraphics[width=\linewidth]{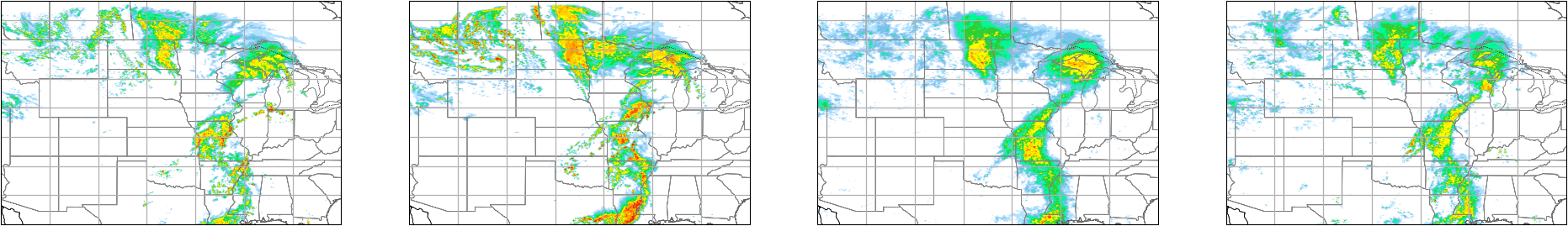} & f15 \\[6pt]
  \includegraphics[width=\linewidth]{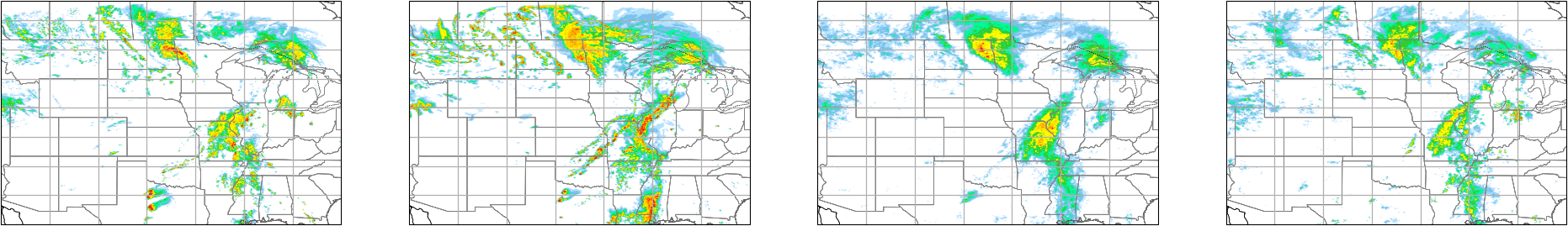} & f18 \\ [6pt]
\multicolumn{2}{c}{
  \begin{minipage}{0.4\linewidth}
    \centering
    \includegraphics[width=\linewidth]{pics/refc-colorbar.png} \\
    \small Composite reflectivity (dBZ)
  \end{minipage}
} \\
\end{tabular}

\caption{Enlarged view of composite reflectivity: forecast initialization time 2024-05-02 06:00 UTC}
\label{refc_map3}

\end{figure}

\begin{figure}
\centering

\noindent
\makebox[0.22\linewidth][c]{HRRR analysis}%
\hfill
\makebox[0.22\linewidth][c]{HRRR forecast}%
\hfill
\makebox[0.22\linewidth][c]{HRRRCast 1-mem.}%
\hfill
\makebox[0.28\linewidth][l]{HRRRCast  determ.}%

\vspace{0.5em}
\includegraphics[width=0.97\linewidth]{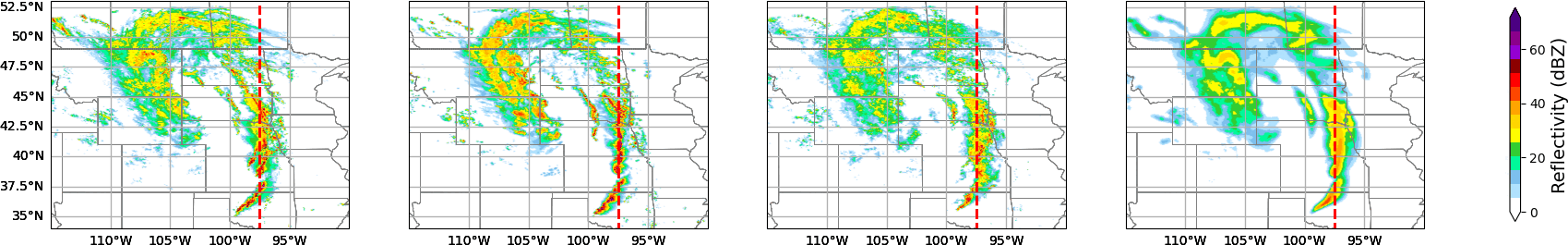}\\
\vspace{0.5em}
\includegraphics[width=0.97\linewidth]{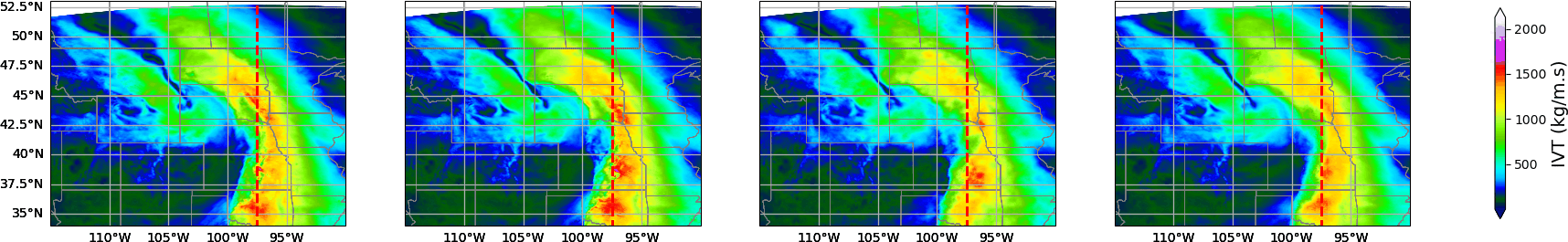} \\
\vspace{0.5em}
\includegraphics[width=0.97\linewidth]{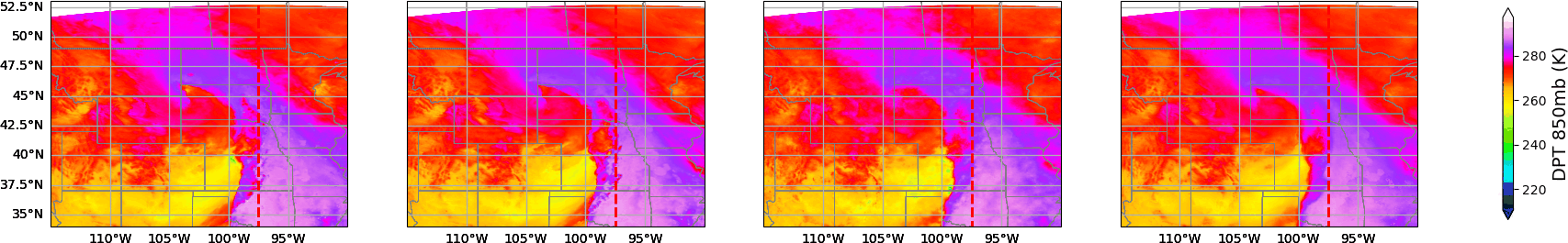} \\
\vspace{0.5em}
\includegraphics[width=1.0\linewidth]{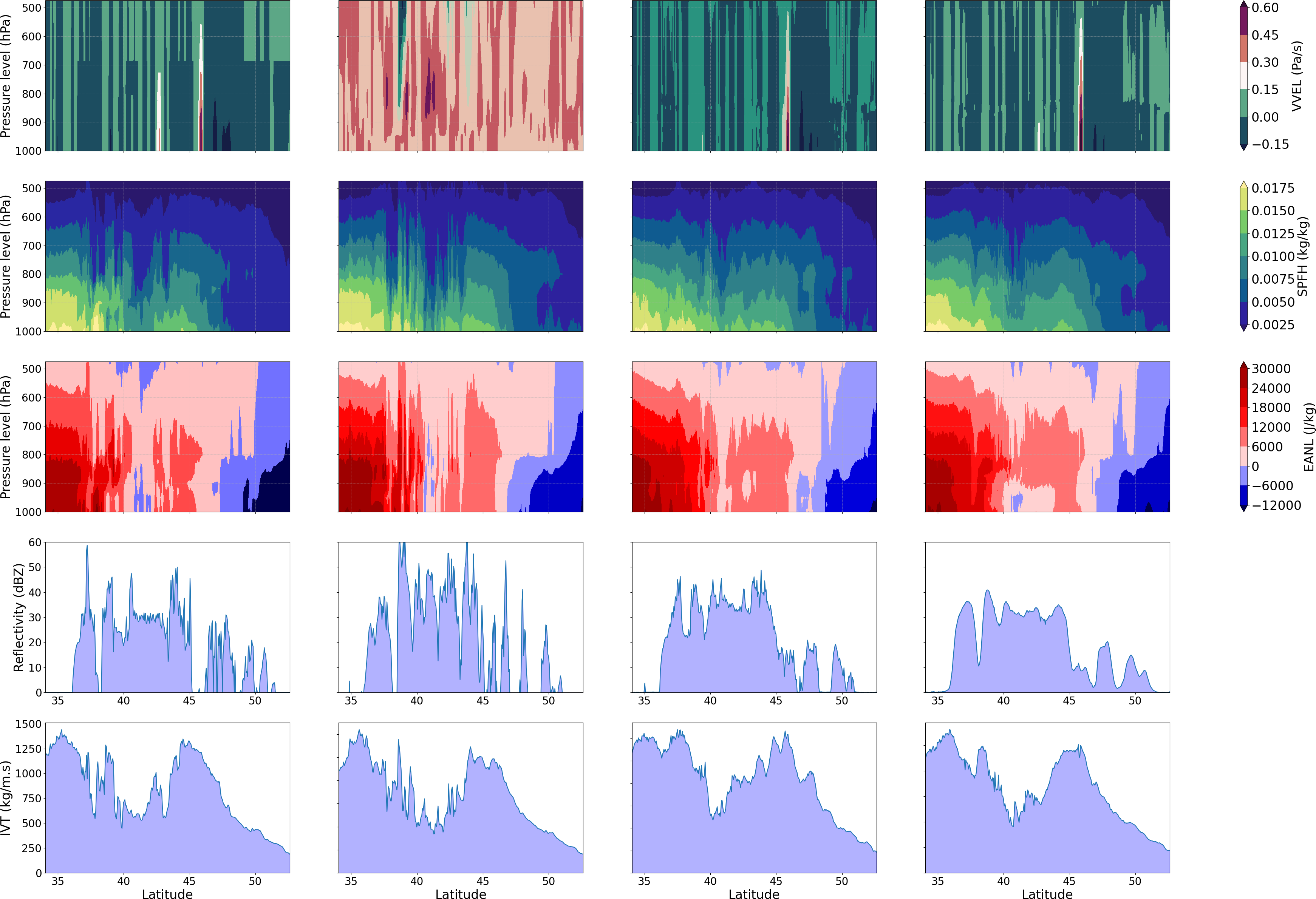} \\

\caption{Case study: A 1h forecast by different models, forecast initialization  time: 2024-05-06 23:00 UTC}
\label{case_study_1h}

\end{figure}

\begin{figure}
\centering

\noindent
\makebox[0.22\linewidth][c]{HRRR analysis}%
\hfill
\makebox[0.22\linewidth][c]{HRRR forecast}%
\hfill
\makebox[0.22\linewidth][c]{HRRRCast 1-mem.}%
\hfill
\makebox[0.28\linewidth][l]{HRRRCast  determ.}%

\vspace{0.5em}
\includegraphics[width=0.97\linewidth]{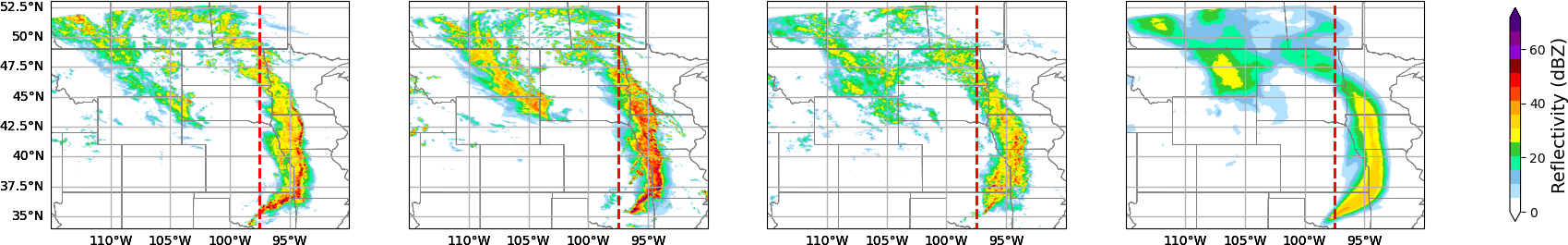}\\
\vspace{0.5em}
\includegraphics[width=0.97\linewidth]{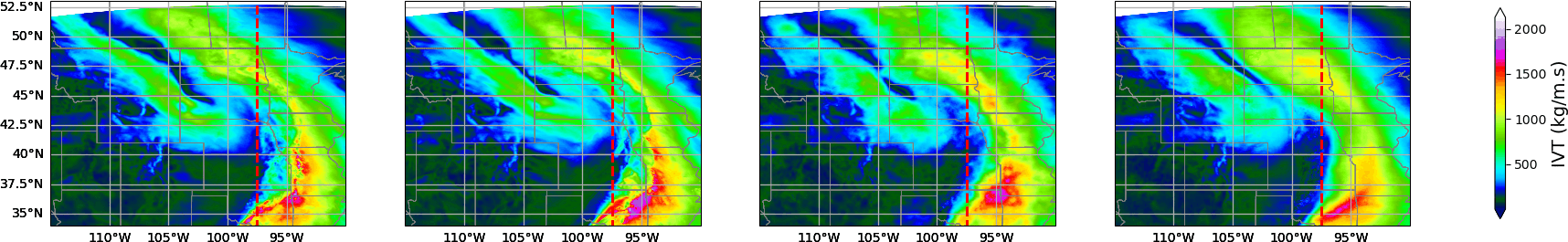} \\
\vspace{0.5em}
\includegraphics[width=0.97\linewidth]{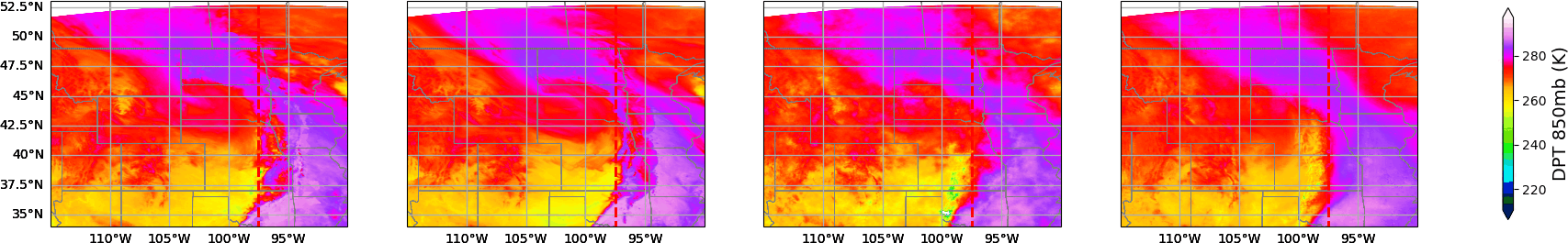} \\
\vspace{0.5em}
\includegraphics[width=1.0\linewidth]{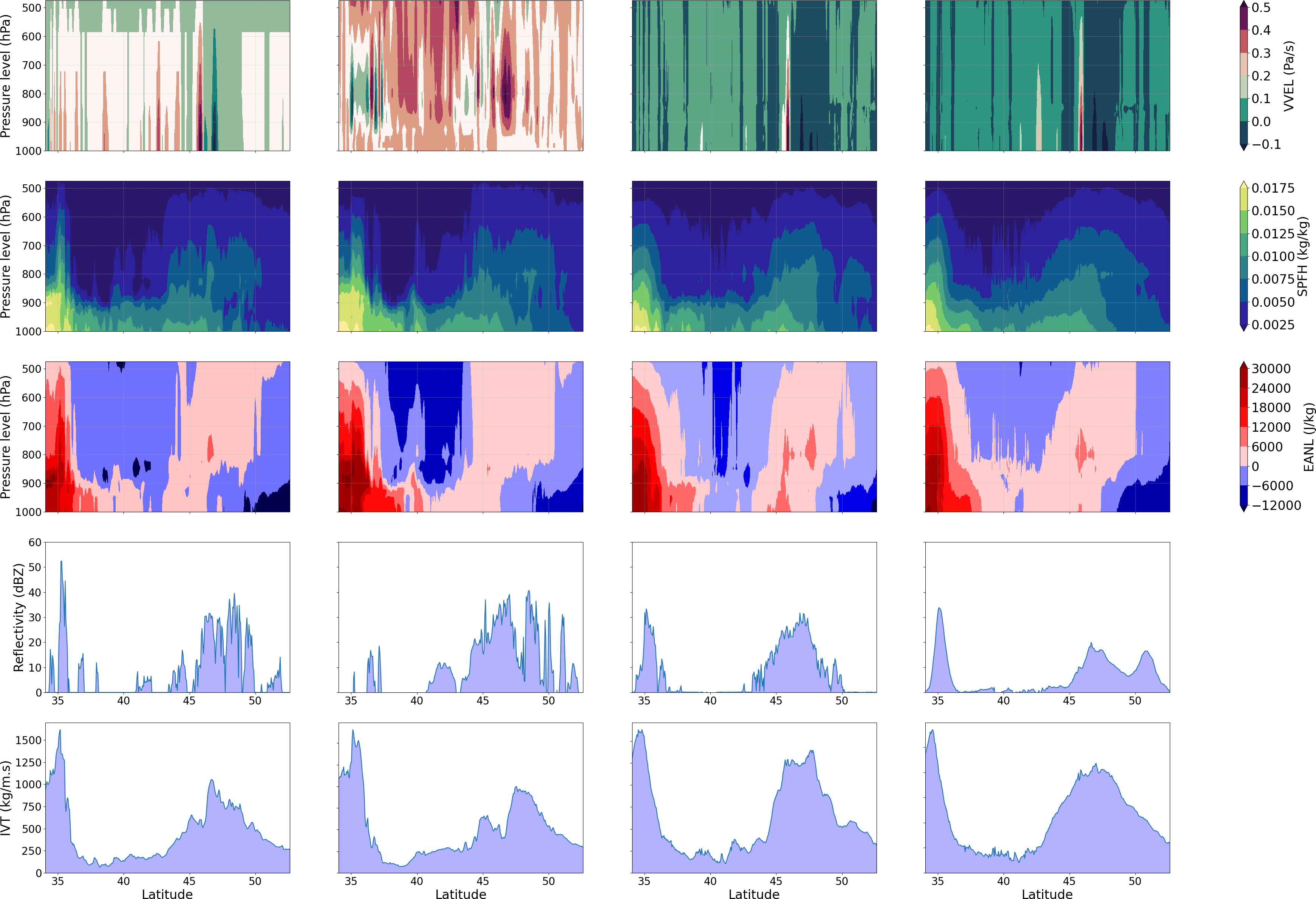} \\

\caption{Case study: A 6h forecast by different models, forecast initialization time: 2024-05-06 23:00 UTC}
\label{case_study_6h}

\end{figure}

\end{document}